\documentclass[tighten,onecolumn]{aastex63}

\usepackage{graphicx}
\usepackage{ulem}
\usepackage{amsmath}
\usepackage{wrapfig}
\usepackage[thinlines]{easytable}
\usepackage{tikz}

\graphicspath{{./}{figures/}}

\begin{document}
\shorttitle{FRBs in nearby globular clusters}
\shortauthors{Kremer et al.}

\title{Prospects for Detecting Fast Radio Bursts in Globular Clusters of Nearby Galaxies}

\correspondingauthor{Kyle Kremer}
\email{kkremer@caltech.edu}

\author[0000-0002-4086-3180]{Kyle Kremer}
\altaffiliation{NSF Astronomy \& Astrophysics Postdoctoral Fellow}
\affiliation{TAPIR, California Institute of Technology, Pasadena, CA 91125, USA}
\affiliation{The Observatories of the Carnegie Institution for Science, Pasadena, CA 91101, USA}

\author[0000-0001-7931-0607]{Dongzi Li}
\affiliation{TAPIR, California Institute of Technology, Pasadena, CA 91125, USA}

\author[0000-0002-1568-7461]{Wenbin Lu}
\affiliation{Departments of Astronomy, Theoretical Astrophysics Center, UC Berkeley, Berkeley, CA 94720, USA}

\author[0000-0001-6806-0673]{Anthony L. Piro}
\affiliation{The Observatories of the Carnegie Institution for Science, Pasadena, CA 91101, USA}

\author[0000-0002-9725-2524]{Bing Zhang}
\affiliation{Nevada Center for Astrophysics, University of Nevada Las Vegas, NV 89154, USA}
\affiliation{Department of Physics and Astronomy, University of Nevada Las Vegas, NV 89154, USA}

\begin{abstract}
The recent detection of a repeating fast radio burst (FRB) in an old globular cluster in M81 challenges traditional FRB formation mechanisms based on magnetic activity in young neutron stars formed
in core-collapse supernovae. Furthermore, the detection of this repeater in such a nearby galaxy implies a high local universe rate of similar events in globular clusters. Building off the properties inferred from the M81 FRB, we predict the number of FRB sources in nearby ($d \lesssim 20\,$Mpc) galaxies with large globular cluster systems known. Incorporating the uncertain burst energy distribution, we estimate the rate of bursts detectable in these galaxies by radio instruments such as FAST and MeerKat. Of all local galaxies, we find M87 is the best candidate for FRB detections. We predict M87's globular cluster system contains $\mathcal{O}(10)$ FRB sources at present and that a dedicated radio survey (by either FAST or MeerKat) of $\mathcal{O}(10)\,$hr has a $90\%$ probability of detecting a globular cluster FRB in M87. The detection of even a handful of additional globular cluster FRBs would provide invaluable constraints on FRB mechanisms and population properties.

Previous studies have demonstrated young neutron stars formed following collapse of dynamically-formed massive white dwarf binary mergers may provide the most natural mechanism for these bursts. We explore the white dwarf merger scenario using a suite of $N$-body cluster models, focusing in particular on such mergers in M87 clusters. We describe a number of outstanding features of this scenario that in principle may be testable with an ensemble of observed FRBs in nearby globular clusters.
\end{abstract}

\section{Introduction}

Fast radio bursts (FRBs) are short, bright flares of coherent radio emission \citep{Lorimer2007,Keane2012,Thornton2013,CordesChatterjee2019} whose origins remain generally unclear \citep[e.g.,][]{Platts2019}. The recent discovery of FRB200428 in association with the Galactic magnetar SGR1935+2154 \citep{CHIME2020,Bochenek2020,Mereghetti2020}
provided key evidence that at least a fraction of FRBs are powered by highly-magnetized neutron stars. Indeed, the magnetar engine model \citep[e.g.,][]{ThompsonDuncan1995,PopovPostnov2013,Kulkarni2014,Lyubarsky2014,Katz2016,Metzger2017,LuKumar18_magnetar_model} can naturally account for many observed properties of FRBs at large, including their short timescales, large energies, repetition \citep{Spitler2016}, and -- since magnetars are traditionally expected to form in association with massive stellar evolution \citep[e.g.,][]{KaspiBeloborodov2017} --
the association of many FRBs with star-forming host galaxies \citep[e.g.,][]{Tendulkar2017,Bhandari2020,Heintz2020,LiZhang2020,Bochenek2021}. As a result, the magnetar model has emerged as perhaps the most popular FRB mechanism model to date.

However, the magnetar model still faces its share of challenges \citep[e.g.,][]{Sridhar2021}. For instance, no presently-confirmed magnetars \citep[e.g.,][]{KaspiBeloborodov2017} are sufficiently active to explain the population of repeaters discovered by CHIME \citep{CHIME2019} which may contribute a signification fraction of the total FRB rate \citep{Margalit2020, Lu20_magnetar_model}.\footnote{Although, given the small Galactic magnetar sample it is possible that a high-activity-level source is simply not expected in the Milky Way, especially if the number of sources varies with activity level as a steep power law.} Also, two of the most well-studied repeaters, FRB 180916 and FRB 121102 exhibit burst periodicity \citep{CHIME2020b, Rajwade2020}. No known magnetars exhibit properties -- for example, binarity \citep[e.g.,][]{Lyutikov2020}, precession \citep[e.g.,][]{Levin2020}, or ultralong spin periods \citep[e.g.,][]{Beniamini2020} -- that have been speculated to lead to periodicity. In addition to its periodicity, FRB 180196 is spatially offset from its host galaxy's closest region of active star formation \citep{Tendulkar2021}, in apparent tension with burst scenarios invoking young magnetars formed in the last $10^3\,$yr via core collapse supernovae (CCSN).

Further straining the classic CCSN magnetar scenario is FRB 20200120E \citep{Majid2021,Bhardwaj2021}, a repeating FRB localized to an old ($t\gtrsim 10\,$Gyr) globular cluster in M81 \citep{Kirsten2022}. Old globular clusters have been devoid of massive star formation
for billions of years, clearly ruling out a young magnetar born through massive stellar collapse as the source of these repeating bursts. Alternatively, \citet{Kirsten2022,Kremer2021,Lu2022} proposed that young highly-magnetized neutron stars formed recently via accretion-induced collapse \citep[e.g.,][]{Tauris2013} or via massive white dwarf binary merger \citep[e.g.,][]{Kremer2021b} may provide a natural formation mechanism for this repeater.
Recently, \citet{Nimmo2022} presented detections of a ``burst storm'' from the M81 repeater providing constraints on the burst energy distribution and wait-time distribution of the source. Additionally, studies have shown this source uniquely exhibits extremely narrow features \citep{Majid2021,Nimmo2022b}. These analyses demonstrate the observed characteristics of the M81 FRB deviate quantitatively from other active repeaters, perhaps hinting at a distinct type of source.

The M81 FRB is also notable as the closest extragalactic FRB observed to date. Simple rate arguments suggest that additional globular-cluster-bound FRBs should be present in the local universe. To this point, \citet{Lu2022} estimated, based on the observed features of the M81 FRB, that a $\sim100\,$hr radio survey of a few nearby ($\lesssim15\,$Mpc) massive galaxies above a fluence threshold of $0.1\rm\, Jy\,ms$ is expected to yield $\mathcal{O}(10)$ detected bursts, a fraction of which may presumably be localized to potential host star clusters similar to the M81 source.

Over the past several decades, our understanding of the sizes of globular cluster systems of nearby galaxies has expanded considerably \citep[e.g.,][]{HarrisRacine1979,Harris1991,BrodieStrader2006,Peng2008,Goergiev2010,Harris2013}, thanks in large part to recent surveys enabled by the Hubble Space Telescope \citep[e.g.,][]{Larsen2001_catalog,Lotz2004,ACSVirgo_2004,ACSFornax_2007, Goergiev2008,Harris2006a,Harris2009}. It is now well-established that old globular clusters are common features of essentially all galaxy types, from ultra-faint dwarfs \citep[e.g.,][]{Crnojevic2016,vanDokkum2017,Danieli2022} to spirals \citep[e.g.,][]{Harris1996,Galleti2004} to massive ellipticals \citep[e.g.,][]{Jordan2009}. This boon in observations has been complemented by a boon in computational methods -- current state-of-the-art simulations are able to perform star-by-star realizations of clusters with up to a million stars over their full lifetimes incorporating not only $N$-body dynamical processes but also stellar evolution \citep[e.g.,][]{Aarseth2003,Giersz2013,wang2016dragon,Rodriguez2022}. The nexus of observations and simulations has enabled detailed studies of various astrophysical phenomena expected in globular clusters in the local universe including X-ray sources \citep[e.g.,][]{Heinke2005,Strader2012,Kremer2018}, millisecond pulsars \citep[e.g.,][]{Ransom2008,Ye2019}, gravitational wave events \citep[e.g.,][]{Rodriguez2016,Askar2017}, and, most recently, FRBs \citep{Kremer2021}.

In this paper, we explore prospects for detecting FRBs similar to the M81 repeater in the globular cluster systems of local galaxies. The paper is organized as follows. In Section~\ref{sec:estimating}, we predict the number of FRB sources and burst detection rates in nearby galaxies. Next, we discuss the specific scenario where the FRBs are powered by young neutron stars born recently through massive white dwarf mergers. In Section~\ref{sec:WD}, we discuss rates and properties of white dwarf mergers using a suite of $N$-body cluster simulations and extrapolate these models to make relevant predictions for M87 globular clusters. In Section~\ref{sec:tests}, we propose several observable tests that may hint at or potentially rule out the white dwarf merger scenario. We summarize our results and conclude in Section~\ref{sec:discussion}.

\section{Estimating local fast radio burst source counts and rates}
\label{sec:estimating}

\begin{deluxetable*}{l|c|c|c|c|c|c|c}
\tabletypesize{\footnotesize}
\tablewidth{0pt}
\tablecaption{Predicted number of FRB sources in galaxies within $20\,$Mpc that are shown in the \citet{Harris2013} catalog to host at least 200 globular clusters. In columns 2-6 we list galaxy properties pulled directly from \citet{Harris2013}. In column 7, we list the inferred number of detectable FRB sources in each galaxy, based on the total number of globular clusters and the inferred number of sources per cluster from the M81 FRB detection. In column 8, we list the predicted white dwarf merger rate in each galaxy's globular cluster system, computed as in Section~\ref{sec:WD}. \label{table:rates}}
\tablehead{
	\colhead{$^1$Galaxy} &
	\colhead{$^2$Dec.} &
	\colhead{$^3M_{\rm v}$} &
	\colhead{$^4$Morph. type} &
	\colhead{$^5d$} &
	\colhead{$^6N_{\rm cl}$} &
	\colhead{$^7N_{\rm src, obs}$} &
	\colhead{$^8$WD merger rate}\\
	\colhead{} &
	\colhead{} &
	\colhead{} &
	\colhead{} &
	\colhead{(Mpc)} &
	\colhead{} &
	\colhead{} &
	\colhead{($\times 10^{-7} \rm{yr}^{-1}$)}
}
\startdata
Milky Way & n/a & -20.6 & Sbc & n/a & 200 & $0.22^{+0.44}_{-0.21}$ & 4.0 \\
\hline
NGC 224 (M31)       & +41 16 08.0 & -21.8 & Sb         & 0.77 & 450 & $0.5^{+0.99}_{-0.47}$ & 9.0 \\
\hline
NGC 3031 (M81)      & +69 03 55.6 & -21.08 & Sab        & 3.55 & 300 & $0.33^{+0.66}_{-0.31}$ & 6.0 \\
\hline
NGC 5128 (Cen A)      & -43 01 05.2 & -21.26 & E3p        & 3.8 & 1300 & $1.43^{+2.86}_{-1.36}$ & 26.0 \\
\hline
NGC 5194 (M51a)      & +47 11 42.5 & -21.16 & Sbc        & 7.62 & 220 & $0.24^{+0.48}_{-0.23}$ & 4.4 \\
\hline
NGC 4517       & +00 06 52.6 & -19.52 & Sc         & 9.26 & 270 & $0.3^{+0.59}_{-0.28}$ & 5.4 \\
\hline
NGC 4594 (M104)      & -11 37 22.8 & -22.12 & Sa         & 9.77 & 1900 & $2.09^{+4.18}_{-1.98}$ & 38.0 \\
\hline
NGC 3115       & -07 43 06.8 & -21.19 & S0         & 10.0 & 550 & $0.6^{+1.21}_{-0.57}$ & 11.0 \\
\hline
NGC 3379       & +12 34 53.8 & -20.9 & E1         & 10.2 & 216 & $0.24^{+0.48}_{-0.23}$ & 4.32 \\
\hline
NGC 1023       & +39 03 47.7 & -21.14 & SB0        & 11.43 & 490 & $0.54^{+1.08}_{-0.51}$ & 9.8 \\
\hline
NGC 4697       & -05 48 02.2 & -21.27 & E6         & 12.01 & 229 & $0.25^{+0.5}_{-0.24}$ & 4.58 \\
\hline
NGC 3556       & +55 40 26.9 & -20.5 & SBc        & 12.42 & 290 & $0.32^{+0.64}_{-0.3}$ & 5.8 \\
\hline
NGC 4636       & +02 41 15.4 & -21.43 & E0         & 14.66 & 4200 & $4.62^{+9.24}_{-4.38}$ & 84.0 \\
\hline
NGC 4621       & +11 38 50.3 & -21.34 & E5         & 14.85 & 803 & $0.88^{+1.77}_{-0.84}$ & 16.06 \\
\hline
NGC 4660       & +11 11 25.9 & -19.74 & E          & 14.97 & 205 & $0.23^{+0.45}_{-0.21}$ & 4.1 \\
\hline
NGC 7331       & +34 24 56.2 & -21.71 & Sbc        & 15.07 & 210 & $0.23^{+0.46}_{-0.22}$ & 4.2 \\
\hline
NGC 4473       & +13 25 45.8 & -20.81 & E5         & 15.25 & 376 & $0.41^{+0.83}_{-0.39}$ & 7.52 \\
\hline
NGC 5866       & +55 45 47.9 & -21.08 & S0-a       & 15.35 & 370 & $0.41^{+0.81}_{-0.39}$ & 7.4 \\
\hline
NGC 4564       & +11 26 21.8 & -20.0 & E          & 15.87 & 213 & $0.23^{+0.47}_{-0.22}$ & 4.26 \\
\hline
NGC 4552       & +12 33 22.0 & -20.39 & E          & 15.89 & 1100 & $1.21^{+2.42}_{-1.15}$ & 22.0 \\
\hline
NGC 4216       & +13 08 58.5 & -21.81 & Sb         & 16.0 & 620 & $0.68^{+1.36}_{-0.65}$ & 12.4 \\
\hline
NGC 4459       & +13 58 42.8 & -20.81 & S0         & 16.01 & 218 & $0.24^{+0.48}_{-0.23}$ & 4.36 \\
\hline
NGC 4278       & +29 16 50.3 & -20.97 & E          & 16.07 & 1100 & $1.21^{+2.42}_{-1.15}$ & 22.0 \\
\hline
NGC 4435       & +13 04 44.1 & -20.4 & SB0        & 16.65 & 345 & $0.38^{+0.76}_{-0.36}$ & 6.9 \\
\hline
NGC 4526       & +07 41 57.1 & -21.51 & S0         & 16.9 & 388 & $0.43^{+0.85}_{-0.41}$ & 7.76 \\
\hline
NGC 4486 (M87)      & +12 23 28.4 & -22.61 & E0         & 17.0 & 13000 & $14.3^{+28.6}_{-13.57}$ & 260.0 \\
\hline
NGC 4472 (M49)      & +08 00 01.4 & -22.82 & E2         & 17.03 & 7000 & $7.7^{+15.4}_{-7.31}$ & 140.0 \\
\hline
NGC 4494       & +25 46 28.8 & -21.4 & E1         & 17.06 & 392 & $0.43^{+0.86}_{-0.41}$ & 7.84 \\
\hline
NGC 4406 (M86)      & +12 56 46.0 & -22.36 & E3         & 17.09 & 2800 & $3.08^{+6.16}_{-2.92}$ & 56.0 \\
\hline
NGC 4649 (M60)      & +11 33 09.6 & -22.41 & E2         & 17.09 & 4000 & $4.4^{+8.8}_{-4.18}$ & 80.0 \\
\hline
NGC 4565       & +25 59 13.9 & -21.68 & Sb         & 17.46 & 204 & $0.22^{+0.45}_{-0.21}$ & 4.08 \\
\hline
NGC 4382 (M85)      & +18 11 26.7 & -22.25 & S0         & 17.88 & 1110 & $1.22^{+2.44}_{-1.16}$ & 22.2 \\
\hline
NGC 4374 (M84)       & +12 53 13.5 & -22.36 & E1         & 18.51 & 4301 & $4.73^{+9.46}_{-4.49}$ & 86.02 \\
\hline
NGC 1553       & -55 46 48.1 & -21.98 & S0         & 18.54 & 540 & $0.59^{+1.19}_{-0.56}$ & 10.8 \\
\hline
NGC 1336       & -35 42 49.2 & -19.14 & E4         & 18.74 & 276 & $0.3^{+0.61}_{-0.29}$ & 5.52 \\
\hline
NGC 1380       & -34 58 34.1 & -21.51 & S0-a       & 18.86 & 424 & $0.47^{+0.93}_{-0.44}$ & 8.48 \\
\hline
NGC 1055       & +00 26 35.4 & -21.3 & Sb         & 19.0 & 210 & $0.23^{+0.46}_{-0.22}$ & 4.2 \\
\hline
NGC 1052       & -08 15 20.9 & -21.05 & E4         & 19.35 & 400 & $0.44^{+0.88}_{-0.42}$ & 8.0 \\
\hline
NGC 1374       & -35 13 34.8 & -20.43 & E          & 19.64 & 360 & $0.4^{+0.79}_{-0.38}$ & 7.2 \\
\hline
NGC 1387       & -35 30 23.9 & -20.84 & E/S0       & 19.82 & 390 & $0.43^{+0.86}_{-0.41}$ & 7.8 \\
\hline
\enddata
\end{deluxetable*}

To estimate FRB demographics in the globular clusters of nearby galaxies, we use the ``Globular Cluster Systems of Galaxies Catalog" \citep{Harris2013}, a compilation of published measurements of known globular cluster populations in 422 local galaxies. This catalog does not contain a complete list of galaxies out to a particular distance \citep[for a more complete galaxy catalog, see e.g.,][]{Tully1988}. However, since the efficiency of globular cluster formation is not strongly dependent on galaxy mass \citep[e.g.,][]{Peng2008,Harris2013} -- except perhaps at the extremes of the mass distribution -- the majority of globular clusters are found in galaxies in a relatively narrow range around $L^{\star}$ \citep{Schechter1976}. Thus, the incompleteness of the \citet{Harris2013} catalog does not strongly affect any of the main conclusions of this paper. Specifically, massive galaxies with large numbers of globular clusters are clearly the best places to look for FRBs in globular clusters. Nonetheless, we note that implementation of a more complete galaxy catalog may have a minor effect on some results. For example, the actual local FRB rate inferred from the M81 detection might be slightly lower (at a percent level) than estimated here since the true number of globular clusters out to the distance of M81 is slightly higher than the number listed in the following paragraph. With these caveats in mind, henceforth we treat the \citet{Harris2013} catalog as ``complete.''

In galaxies out to the distance of M81 \citep[$d\approx3.6\,$Mpc;][]{Karachentsev2005} that are observable by CHIME \citep[above roughly $-20^{\circ}$ declination;][]{CHIME2018}, there are estimated to be roughly 900 total globular clusters \citep{Harris2013}, including the roughly 50 Galactic clusters with sufficiently high declinations \citep{Harris1996}. The fluence of the bursts detected by CHIME \citep[$\approx2\,\rm{Jy\,ms}$;][]{Bhardwaj2021} from the M81 source are roughly right at the CHIME fluence detection threshold. After M81, the next closest galaxy with a large known globular cluster system (above $-20^{\circ}$) is about two times further away \citep[NGC~5457 -- or M101 -- with about 100 clusters;][]{Harris2013}. To increase the enclosed number of observable clusters by an order unity factor, one would need to go out to NGC~4594 at $d\approx10\,$Mpc. At this distance, the $2\,\rm{Jy\,ms}$ M81 bursts detected by CHIME would be significantly below CHIME’s fluence threshold. In this case, the detection of a single FRB repeater by CHIME in the globular clusters enclosed out to M81 implies a specific abundance of $N_{\rm obs}' \approx 1/900 \approx1.1\times10^{-3}$ observed FRB sources per globular cluster. Incorporating the Poisson probability associated with a single event \citep[as discussed in][]{Lu2022}, this suggests a specific abundance range of $N_{\rm obs}' \in (5.6\times10^{-5},3.3\times10^{-3})$ at $90\%$ confidence. The number of observable sources in a given galaxy with $N_{\rm cl}$ total globular clusters can then be estimated simply as $N_{\rm obs, gal} \approx N_{\rm cl} \times N_{\rm obs}'$. Taking into account a beaming factor, $f_b$, the true total number of sources is  $N_{\rm gal} \approx N_{\rm obs, gal}\,f_b^{-1}$. In Table \ref{table:rates}, we list all galaxies within $20\,$Mpc predicted to host at least $200$ globular clusters \citep{Harris2013}. In column 7, we list the predicted number of observable FRB sources in each galaxy's globular cluster system (with $90\%$ confidence interval ranges as computed from the Poisson distribution as discussed previously). In instances where the predicted number of sources is less than unity, the value of column 7 can be interpreted as the probability of the given galaxy containing a single observable source. 

Given a predicted number of FRB sources, we can next predict the detectable burst rate per galaxy. Of course, this rate is only meaningful if at least one FRB-emitting source is present. Therefore, we present this estimate only for the galaxies in Table \ref{table:rates} predicted to contain at least one source. For a telescope operating at fluence completeness threshold, $F_{\nu,{\rm th}}$, the telescope is sensitive to bursts above an energy threshold $E_{\rm th} =4\pi^2 D^2 \nu F_{\nu,{\rm th}}$. In this case, the detectable burst rate at a given distance is determined by the assumed burst energy distribution, which is expected to take the form $dN/dE \propto E^\alpha$ with the uncertain power-law index $\alpha$ potentially varying from roughly $-1$ to $-4.5$ \citep[e.g.,][]{Luo2018,LuPiro19_FRB_rate_density,Hashimoto2020,Cruces2021,Lanman2022,Nimmo2022}. The number of bursts above $E_{\rm th}$ at some distance $D$ then takes the form $N \propto (D^2F_{\nu, \rm{th}})^{\alpha + 1}$. CHIME detected 7 bursts from the M81 source over an on-source time of roughly $100\,$hr, implying a burst rate of roughly $0.07\,\rm{hr}^{-1}$. The burst rate expected for bursts detectable by an arbitrary radio telescope of fluence threshold, $F_{\nu, \rm{th}}$, at some distance, $D$, can then be written as

\begin{deluxetable*}{c|c|c||l|c|c|c}
\tabletypesize{\footnotesize}
\tablewidth{0pt}
\tablecaption{Predicted FRB detection rate in nearby galaxies with at least one observable FRB source predicted at present. Where relevant (based on the galaxy declination values), we show detection rates for both FAST and MeerKat. See text for detailed discussion. Column 7 ($\alpha=-2.4$; in bold) should be viewed as our fiducial rate estimate since this adopts the inferred burst energy distribution of \citet{Nimmo2022}. Columns 5 and 6 show rates for more pessimistic burst energy distributions. The rate ranges shown in parentheses show the $90\%$ confidence range inferred from adopting a Poisson distribution, as described in the text. \label{table:FRBrates}}
\tablehead{
	\colhead{} &
	\colhead{} &
	\colhead{} &
	\colhead{} &
	\multicolumn{3}{c}{FRB rate $[\rm{hr}^{-1}]$}\\
	\colhead{$^1$Galaxy} &
	\colhead{$^2d$ [Mpc]} &
	\colhead{$^3r_{\rm{cl},50}$ [arcmin]} &
	\colhead{$^4$Telescope} &
	\colhead{$^5\alpha=-1.5$} &
	\colhead{$^6\alpha=-2$} &
	\colhead{$^7\bf{\alpha=-2.4}$}
}
\startdata
NGC 5128       & 3.8 & $37.2^i$ & FAST: & N/A & N/A & N/A \\
(Cen A) & & & MeerKat: & $0.33\,(0.02-0.98)$ & $2.3\,(0.12-6.89)$ & $\bf{10.97\,(0.56-32.9)}$ \\
\hline
NGC 4594       & 9.77 & $6.0^d$ & FAST: & $0.22\,(0.01-0.66)$ & $1.47\,(0.07-4.41)$ & $\bf{6.75\,(0.34-20.26)}$ \\
(M104) & & & MeerKat: & $0.4\,(0.02-1.21)$ & $1.1\,(0.06-3.31)$ & $\bf{2.48\,(0.13-7.43)}$ \\
\hline
NGC 4636       & 14.66 & $4.5^a$ & FAST: & $0.43\,(0.02-1.28)$ & $1.91\,(0.1-5.73)$ & $\bf{6.35\,(0.32-19.04)}$ \\
& & & MeerKat: & $0.59\,(0.03-1.78)$ & $1.08\,(0.06-3.25)$ & $\bf{1.76\,(0.09-5.27)}$ \\
\hline
NGC 4278       & 16.07 & $2.82^b$ & FAST: & $0.1\,(0.01-0.3)$ & $0.41\,(0.02-1.22)$ & $\bf{1.26\,(0.06-3.78)}$ \\
& & & MeerKat: & N/A & N/A & N/A \\
\hline
NGC 4486       & 17.0 & $11.4^c$ & FAST: & $0.59\,(0.03-1.78)$ & $2.29\,(0.12-6.87)$ & $\bf{6.75\,(0.34-20.26)}$ \\
(M87) & & & MeerKat: & $1.45\,(0.07-4.36)$ & $2.29\,(0.12-6.88)$ & $\bf{3.31\,(0.17-9.92)}$ \\
\hline
NGC 4472       & 17.03 & $8.7^d$ & FAST: & $0.38\,(0.02-1.14)$ & $1.47\,(0.07-4.41)$ & $\bf{4.33\,(0.22-12.98)}$ \\
(M49) & & & MeerKat: & $0.85\,(0.04-2.55)$ & $1.34\,(0.07-4.01)$ & $\bf{1.93\,(0.1-5.78)}$ \\
\hline
NGC 4406       & 17.09 & $9.1^e$ & FAST: & $0.15\,(0.01-0.44)$ & $0.56\,(0.03-1.69)$ & $\bf{1.66\,(0.08-4.97)}$ \\
(M86) & & & MeerKat: & $0.34\,(0.02-1.02)$ & $0.53\,(0.03-1.59)$ & $\bf{0.76\,(0.04-2.29)}$ \\
\hline
NGC 4649       & 17.09 & $4.2^f$ & FAST: & $0.36\,(0.02-1.07)$ & $1.37\,(0.07-4.1)$ & $\bf{4.01\,(0.2-12.04)}$ \\
(M60) & & & MeerKat: & $0.48\,(0.02-1.45)$ & $0.76\,(0.04-2.28)$ & $\bf{1.09\,(0.06-3.27)}$ \\
\hline
NGC 4382       & 17.88 & $1.92^g$ & FAST: & $0.09\,(0.0-0.28)$ & $0.35\,(0.02-1.04)$ & $\bf{0.98\,(0.05-2.95)}$ \\
(M85) & & & MeerKat: & N/A & N/A & N/A \\
\hline
NGC 4374       & 18.51 & $2.9^h$ & FAST: & $0.35\,(0.02-1.06)$ & $1.25\,(0.06-3.76)$ & $\bf{3.45\,(0.18-10.36)}$ \\
(M84) & & & MeerKat: & $0.48\,(0.02-1.44)$ & $0.7\,(0.04-2.09)$ & $\bf{0.94\,(0.05-2.81)}$ \\
\hline
\enddata
\tablecomments{References for observed cluster surface density profiles: (a) \citet{Dirsch2005}; (b) \citet{Usher2013}; (c) \citet{Strader2011}; (d) \citet{RhodeZepf2001}; (e) \citet{RhodeZepf2004}; (f) \citet{Lee2008}; (g) \citet{Escudero2022}; (h) \citet{Gomez2004}; (i) \citet{Hughes2021_CenA}}
\end{deluxetable*}

\begin{equation}
    \label{eq:rate}
    R\approx 0.07 \Bigg( \frac{F_{\nu,\rm{th}}}{5\,\rm{Jy\,ms}} \Bigg)^{\alpha+1} \Bigg( \frac{D}{3.6\,\rm{Mpc}} \Bigg)^{2(\alpha+1)}\, \rm{hr}^{-1}
\end{equation}
where we have scaled according to CHIME's detected burst rate at $F_{\nu, \rm{th}}=5\,\rm{Jy\,ms}$. In principle, Equation~(\ref{eq:rate}) should also contain a frequency dependence, however since different repeaters may very well exhibit different frequency dependencies, we simply neglect this dependence here.
Recently, \citet{Nimmo2022} reported 60 bursts detected with the Effelsberg telescope ($F_{\nu, \rm{th}}=0.16\,\rm{Jy}\,{ms}$) for an observing duration of roughly $28\,$hr. As a sanity check, Equation~(\ref{eq:rate}) implies a detectable burst rate of roughly $2-8\,\rm{hr}^{-1}$ for $F_{\nu, \rm{th}}=0.16\,\rm{Jy\,ms}$ and $\alpha\in(-2,-2.4)$, consistent to within a small factor of the detected burst rate from \citet{Nimmo2022}.

In Table \ref{table:FRBrates}, we show the estimated burst detection rate for the ten galaxies shown in Table \ref{table:rates} predicted to host at least one (appropriately beamed) FRB source for two radio telescopes: MeerKat and FAST. The detection rates shown incorporate the relevant fluence threshold for each telescope as well as the fraction of each galaxy's globular clusters that are enclosed within each telescope's field of view. For MeerKat, at $\nu=1.4\,$GHz, we adopt a fluence threshold of $F_{\nu, \rm{th}}=0.09\,\rm{Jy}\,{ms}$ and field of view of $r=31\,$arcmin \citep{Jonas2016,Bailes2020_MeerKat}. For FAST, the full width at half maximum (FWHM) of each of the 19 beams at $\nu=1.3\,$GHz is $3\,$arcmin, indicating a total coverage of roughly $134\,\rm{arcmin}^2$ at fluence threshold $F_{\nu, \rm{th}}=0.015\,\rm{Jy}\,{ms}$ \citep{Jiang2020,Niu2021_FAST}.\footnote{This is the fluence threshold value at the center of the FAST beams. At the beam edges, the sensitivity is lower (by definition roughly $0.03\,\rm{Jy\,ms}$ at FWHM). Thus, the adopted $F_{\nu, \rm{th}}=0.015\,\rm{Jy}\,{ms}$ is slightly optimistic. Adopting a more conservative fluence threshold value of $F_{\nu, \rm{th}}=0.03\,\rm{Jy}\,{ms}$ would yield a predicted detection rate roughly $2-3$ times lower.} The full 19-beam configuration covers a field of view of radius roughly $12\,$arcmin, however the beam centers are each separated by roughly $5.8\,$arcmin \citep{Jiang2020}. Thus, the total coverage of the 19-beam $r=12\,$arcmin field of view is roughly $30\%$. This coverage fraction is incorporated into the FAST rates shown in the table.

In the table, we adopt three possible values for the uncertain power-law  exponent $\alpha$: First, we assume $\alpha=-2.4$ as motivated by the observed luminosity function of the M81 repeater from \citet{Nimmo2022} down to Effelsberg detection threshold $F_{\nu,\rm{th}}=0.16\,$Jy~ms which, for $\nu=1.4\,$GHz and $d=3.6\,$Mpc, corresponds to a minimum detectable energy of roughly $3.5\times10^{33}\,$erg. This is lower than the corresponding minimum detectable energy for $F_{\nu,\rm{th}}=0.015\,$Jy~ms at the distance of M87 (roughly $7.3\times10^{33}\,$erg). Second, since it is unknown whether the energy distribution of the M81 repeater is necessarily representative of all globular cluster FRBs, we also adopt $\alpha=-2$ and $\alpha=-1.5$ as shallower energy distribution cases.

\begin{figure*}
    \centering
    \includegraphics[scale=0.8]{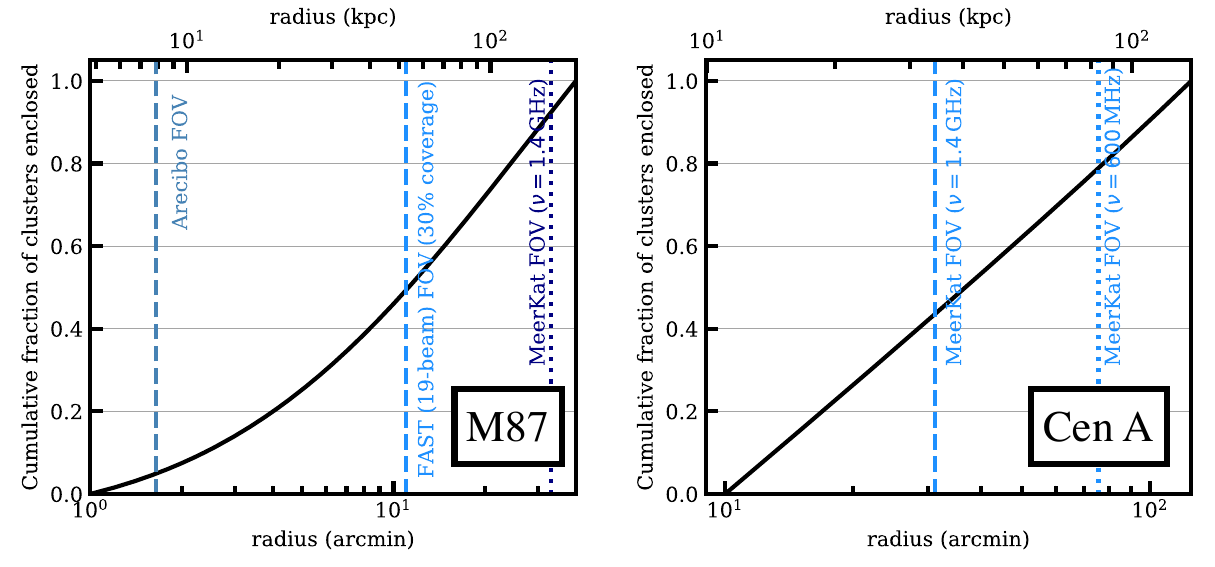}
    \caption{Cumulative fraction of globular clusters versus 2D radial position for M87 and Cen A. For reference, as vertical dashed lines we show the radius covered by the field of view for a few different radio telescopes. As discussed in the text, the $r\approx12\,$arcmin field of view shown for FAST's 19-beam configuration features only $\approx30\%$ coverage, due to the spacing between the beams.}
    \label{fig:radial_dist}
\end{figure*}

In Figure~\ref{fig:radial_dist}, we show the cumulative fraction of globular clusters enclosed versus 2D projected radius for M87 and Cen A, computed from the cluster surface density profiles of \citet{Strader2011} and \citet{Hughes2021_CenA}, respectively. We show as blue vertical lines the fields of view for the 19-beam configuration of FAST \citep{Nan2011_FAST} and for MeerKat at both $\nu=1.4\,$GHz and $600\,$MHz \citep{Jonas2016}. As shown, the respective field of views for FAST and MeerKat (for $\nu=1.4\,$GHz) each span roughly the innermost 50\% of the globular cluster systems of M87 and Cen A (however, FAST features only $\approx30\%$ coverage due to spacing between beams as described above). For reference, in column 3 of Table~\ref{table:FRBrates}, we list the 2D radius that contains half of the globular cluster population, $r_{\rm cl,50}$, for all galaxies listed. The fraction of clusters enclosed within a given radius are computed by integrating the observed cluster surface density profiles for the various galaxies (see references listed in the caption of the table). 

Table~\ref{table:FRBrates} establishes M87 and Cen A as the most promising galaxies for detecting globular cluster FRBs, in the north and south, respectively. Given that M87 has by far the most globular clusters ($N_{\rm cl}\gtrsim10^4$) of any nearby galaxy (see column 6 of Table~\ref{table:rates}), this is not a surprise. Although Cen A contains a factor of roughly $10$ fewer clusters relative to M87, its relative proximity enables lower luminosity bursts to be detected. Based on the $90\%$ lower limits of Table~\ref{table:FRBrates}, we predict that a roughly $30\,$hr ($3\,$hr) FAST survey of M87 has a $90\%$ chance of detecting a single burst, assuming $\alpha=-1.5$ ($\alpha=-2.4$). A MeerKat survey of M87 of roughly  $14\,$hr ($6\,$hr) has a $90\%$ chance of detecting a burst for $\alpha=-1.5$ ($\alpha=-2.4$). Additionally, we predict that a $50\,$hr ($1.8\,$hr) MeerKat survey of Cen A has a $90\%$ chance of detecting a single burst for $\alpha=-1.5$ ($\alpha=-2.4$).

Since the 1980s, M87 has been a popular galaxy target for surveys of pulsed radio emission \citep[e.g.,][]{LinscottErkes1980,McCulloch1981,Taylor1981}, with most cases yielding null results. Recently, \citet{Suresh2021} conducted a $\approx10\,$hr survey of the core of M87 with the Aricebo radio telescope (in the frequency range $1.15-1.75\,$GHz and with minimum detectable fluence $1.4\,\rm{Jy\,ms}$) and identified no evidence of astrophysical bursts. Note this fluence threshold is much greater than the sensitivity limit of the telescope, due to the extremely bright supermassive black hole at the center of M87. Much better sensitivity can be achieved when observing with a multi-beam configuration offset from the center of the galaxy. For the $1.4\,\rm{Jy\,ms}$ fluence threshold, Equation~(\ref{eq:rate}) suggests roughly $0.8$ detectable bursts in M87 for a $10\,$hr observing window (for $\alpha=-2.4)$, assuming the full globular cluster system is monitered. However, as shown in the left-panel of Figure\,\ref{fig:radial_dist}, the half power beam width of $3.3\,$arcmin of this Aricebo observation spans only the innermost $\approx5\%$ of the full globular cluster population. In this case, the lack of detection after $10\,$hr is expected. Wider field of views and lower fluence thresholds (e.g., as enabled by FAST or MeerKat) are necessary to detect high rates of bursts in M87. 

\section{White dwarf binary mergers}
\label{sec:WD}

Until this point, we have remained agnostic toward the specific formation channel through which these globular cluster FRB sources may be formed. Previous studies \citep[e.g.,][]{Kremer2021,Lu2022} have argued that perhaps the most plausible explanation is a young neutron star born from a recent super-Chandrasehkhar white dwarf binary merger. In this section, we discuss the white dwarf merger scenario in more detail. In Section~\ref{sec:tests}, we go onto describe ways this scenario may be tested.

\begin{figure}
    \centering
    \includegraphics[scale=0.64]{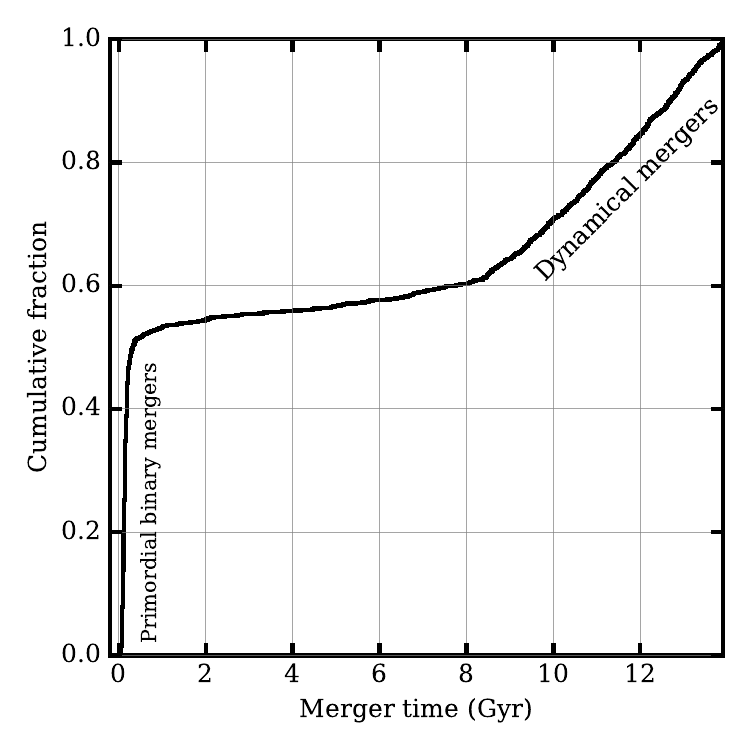}
    \caption{Cumulative fraction of merger times for all white dwarf mergers occurring in our suite of \texttt{CMC} simulations. As described in the text, the peak at early times ($t\lesssim300\,$Myr) arises from primordial binaries that merge following a common envelope episode, while the peak at later times ($\gtrsim9$\,Gyr) arises through dynamical encounters of white dwarfs as described in \citet{Kremer2021b}.}
    \label{fig:delay_time}
\end{figure}

\subsection{Motivation}

It is now well-established that, in globular clusters, compact object populations play a crucial role in the evolution of their host environment \citep[e.g.,][]{Mackey2007,BreenHeggie2013,kremer2020modeling,AntoniniGieles2020}. At early times ($t\lesssim10\,$Gyr), clusters are expected to harbor stellar-mass black hole subsystems in their centers, which dynamically ``heat'' their host cluster through frequent dynamical encounters \citep[e.g.,][]{Kremer2020burning}. As the cluster evolves, the black hole subsystem erodes as black holes are dynamically ejected \citep[e.g.,][]{Kulkarni1993}. At late times ($t\gtrsim10\,$Gyr), once nearly all black holes have been ejected, some clusters can undergo core collapse \citep[e.g.,][]{kremer2019initial} at which point massive white dwarfs mass segregate and form their own dense central subsystem \citep[e.g.,][]{Kremer2021b,Vitral2022}. Within the white dwarf subsystems of core-collapsed clusters, white dwarf binaries form dynamically leading to high rates of white dwarf binary mergers, the vast majority of which have a total mass in excess of the Chandrasekhar limit suggesting many of these mergers lead to collapse and formation of young neutron stars \citep[e.g.,][]{NomotoIben1985,Schwab2016,Schwab2021}. 

Using $N$-body cluster models of the core-collapsed Milky Way globular cluster NGC 6397, \citet{Kremer2021b} predicted a volumetric merger rate of super-Chandrasekhar white dwarf binaries in globular clusters in the local universe of roughly $10\,\rm{Gpc}^{-3}\,\rm{yr}^{-1}$. If all of these white dwarf mergers lead to formation of a young highly-magnetic neutron star capable of emitting FRBs, we obtain an FRB source formation rate of $\dot{n}\approx 10\,\rm{Gpc}^{-3}\,\rm{yr}^{-1}$. As discussed in \citet{Kremer2021,Lu2022}, the detection of a single FRB repeater in M81 implies a volumetric density of observable FRB sources of roughly $n_{\rm obs}\approx 5\times10^6\,\rm{Gpc}^{-3}$ in the local universe or $n_{\rm obs} \in (2.6\times10^5\,\rm{Gpc}^{-3},1.4\times10^7\,\rm{Gpc}^{-3})$ at 90\% confidence, incorporating the Poisson probability associated with detection of a single source. Adopting a beaming factor $f_b\approx0.3$ \citep{Lu2022}, we infer a true source density of $n\approx f_b^{-1}n_{\rm obs} \approx10^7\rm{Gpc}^{-3}$. The inferred source activity timescale, $\tau \approx f_b^{-1}n_{\rm obs}/\dot{n}$, for neutron stars formed from massive white dwarf mergers is then roughly $1.6\times10^6\,$yr or $(8.7\times10^4\,\rm{yr},4.7\times10^6\,\rm{yr})$ at 90\% confidence. Notably, this inferred timescale is consistent with the magnetic activity timescale \citep[e.g.,][]{BeloborodovLi2016} expected for neutron stars formed through collapse following white dwarf mergers \citep[e.g.,][]{Schwab2016,Schwab2021}. Furthermore, the associated magnetic energy budget is consistent with the burst energetics observed for the M81 FRB \citep{Kremer2021,Lu2022}.

With this motivation in mind, in the following subsections, we go on to discuss white dwarf mergers occurring in a broad set of $N$-body cluster simulations. As the globular clusters of M87 seem to be the most promising target for future FRBs (see Table~\ref{table:FRBrates}), we predict white dwarf merger rates and properties for this galaxy.

\subsection{Globular cluster models}

In \citet{Kremer2021b}, we discussed the dynamics of white dwarfs in the specific cluster NGC\,6397 \citep[motivated by this cluster's observationally-inferred white dwarf population;][]{VitralMamon2021,Vitral2022} and used these results to extrapolate the white dwarf merger rate in the local universe. Here we use the \texttt{CMC Cluster Catalog} \citep{kremer2020modeling}, a much more expansive set of $N$-body cluster simulations that encapsulate the full range of globular cluster properties observed in the Milky Way.

\texttt{CMC} \citep{Rodriguez2022}, is a H\'{e}non-type \citep{Henon1971} Monte Carlo $N$-body code that includes relevant physical processes for modeling compact objects in globular clusters, including two-body relaxation, direct integration of small-$N$ resonant encounters, treatment of stellar mergers/tidal disruptions, tidal stripping in a galactic potential, and stellar/binary star evolution \citep[using \texttt{COSMIC};][]{Breivik2020}. The latest grid of models, the \texttt{CMC Cluster Catalog}, includes roughly $150$ independent simulations with initial $N$ ranging from $2\times10^5-3.6\times10^6$, initial virial radii, $r_v$, ranging from $0.5-4\,$pc, metallicity ranging from $0.01-1\,Z_{\odot}$, and positions in the Galactic potential ranging from $2-20\,$kpc. As described in \citet{kremer2020modeling}, this catalog of models as a whole effectively covers the full parameter space of interest of the globular clusters observed in the Milky Way and enables detailed study of various phenomena pertaining to compact objects in clusters including formation of X-ray binaries, millisecond pulsars, AM CVn, and gravitational-wave sources.

\begin{figure*}
    \centering
    \includegraphics[scale=0.7]{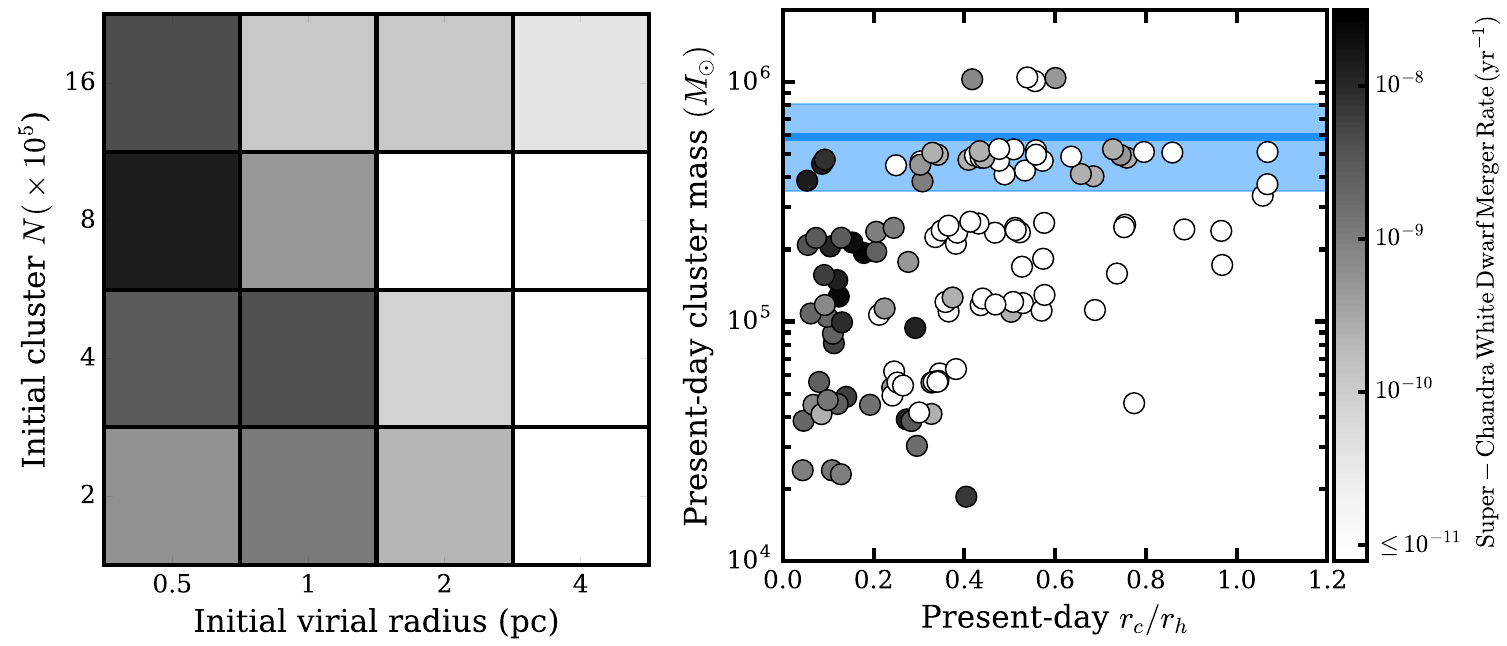}
    \caption{Merger rate of super-Chandrasekhar white dwarfs at late times for our assumed grid of cluster initial properties (left) and properties of cluster models at present-day ages (right). In blue we show the inferred mass of the globular cluster, [PR95] 30244,  hosting the M81 FRB \citep[from][]{Kirsten2022}.}
    \label{fig:rate_density}
\end{figure*}

\newpage
\subsection{Rates and dependence on cluster properties}

In total, 2451 white dwarf mergers occur in the suite of 148 \texttt{CMC catalog} models. Of these, 875 (1576) occur at late times, $t>9\,$Gyr (early times; $t<9\,$Gyr). Of the early time mergers, roughly 75\% occur at $t<300\,$Myr -- these occur through following common envelope evolution of primordial binaries of component masses $\approx5\,M_{\odot}$ \citep[e.g.,][]{Webbink1984}. Mergers occurring at late times are formed exclusively through dynamical encounters \citep[for discussion of the relevant dynamics, see][]{Kremer2021b}. For the purposes of creating FRB sources in old globular clusters, the late-time dynamical mergers are of relevance; neutron stars created through collapse following the primordial binary mergers at early times will have long since undergone spin down and magnetic field decay, rendering them undetectable as FRB sources at the present day. In Figure~\ref{fig:delay_time}, we show the distribution (cumulative fraction) of merger times for all white dwarf mergers occurring in the simulations.

The rate of these late time mergers in old globular clusters depends sensitively on the host clusters' properties, specifically the cluster density. In Figure~\ref{fig:rate_density}, we show the white dwarf merger rate per cluster (limited to late times; $t>9\,$Gyr) for different cluster properties. In the left panel, we show the rate as a function of initial properties: the initial number of objects in the cluster, $N$, versus the initial virial radius, $r_v$. Both of these values are set as initial conditions in the $N$-body simulations \citep[see][]{kremer2020modeling}.
As shown, the merger rate is highest for models with $N=8\times10^5$ and $r_v=0.5\,$pc, which can be explained as follows. The total white dwarf merger rate in a cluster scales as $\Gamma \propto N_{\rm WD} n_{\rm WD} \Sigma \sigma_v$. $N_{\rm WD}$ is the number of white dwarfs in the cluster, $n_{\rm WD}$ is the number density of white dwarfs, $\Sigma$ is the relevant cross section for mergers, and $\sigma_v$ is the cluster's central velocity dispersion. $N_{\rm WD}$ and $n_{\rm WD}$ play the primary role.\footnote{$\Sigma$ is determined by white dwarf radii which can be considered constant across cluster types. Additionally, $\sigma_v\propto N^{1/2}$ varies by less than a factor of roughly 10 across cluster types.} $N_{\rm WD}$ is determined simply by the total number of stars in the cluster, $N$; all other things being equal, more massive clusters feature more white dwarfs and thus more white dwarf mergers. $n_{\rm WD}$ can vary by several orders of magnitude depending on cluster properties. In non-core-collapsed clusters, $n_{\rm WD}$ is likely comparable to the overall cluster density, typically $10^3-10^4\,\rm{pc}^{-3}$. However, in core-collapsed clusters where the white dwarfs are expected to have mass-segregated and formed a dense white dwarf subsystem, $n_{\rm WD}$ can reach as high as $10^6\,\rm{pc}^{-3}$ \citep[e.g.,][]{Kremer2021b}. The time to reach core-collapse is determined by the cluster's half-mass relaxation timescale \citep[e.g.,][]{Spitzer1987}

\begin{equation}
    t_{\rm relax} \propto \frac{N^{1/2}}{\langle m \rangle^{1/2} \ln{N}} r_v^{3/2}.
\end{equation}
Due to their relatively short relaxation times, clusters with smaller $r_v$ are more likely to have reached core collapse by the present day and therefore feature more white dwarf mergers, as clearly evidenced by the $r_v=0.5\,$pc models in Figure~\ref{fig:rate_density}. For a fixed $r_v$, $t_{\rm relax}$ increases with $N$. In our case, the $N=1.6\times10^6,r_v=0.5\,$pc models have a sufficiently long relaxation time to have not yet reached core-collapse. As a result, these models feature fewer mergers, despite the fact that $N_{\rm WD}$ has increased.

In the right hand panel of Figure~\ref{fig:rate_density}, we show merger rates for late-time cluster properties. As clusters evolve, $r_c$ decreases and $r_h$ increases \citep[as energy flows from the cluster's core to its halo; see, e.g.,][]{HeggieHut2003}. This panel again demonstrates the clear overabundance of late-time white dwarf mergers in core-collapsed clusters -- the most dynamically evolved clusters with the lowest $r_c/r_h$ values. For reference, we show as a blue band the inferred mass of the host globular cluster of the M81 FRB \citep[see discussion in][]{Kirsten2022}.

\begin{figure*}
    \centering
    \includegraphics[scale=0.8]{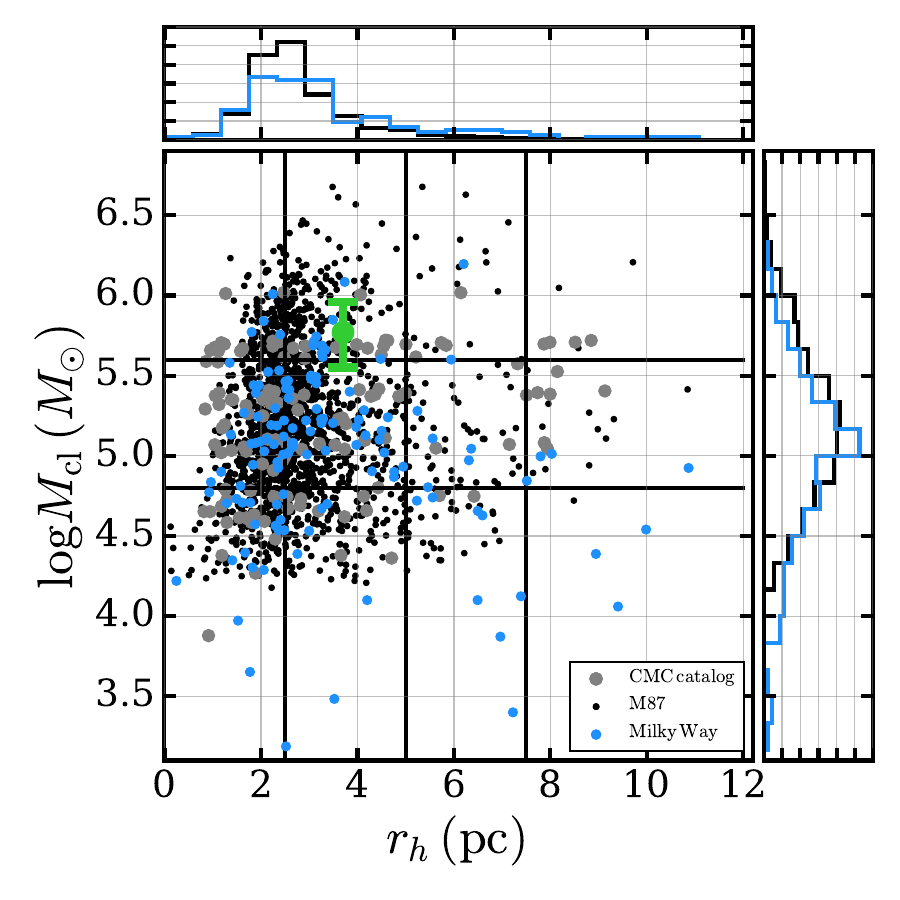}
    \caption{Cluster properties for Milky Way (blue) and M87 (black) clusters compared to CMC catalog models (gray). As a green scatter point, we show the inferred properties of the host globular cluster of the M81 FRB. Solid black lines denote the boundaries adopted for our weighting scheme used to translate the \texttt{CMC} models to an M87 sample, as described in Section~\ref{sec:scaling}.}
    \label{fig:CMCmodels}
\end{figure*}

\subsection{Scaling the \texttt{CMC Catalog} to M87}
\label{sec:scaling}

The \texttt{CMC Cluster Catalog} models were originally intended as a proxy for the properties and total number of Milky Way globular clusters. The key difference between the globular cluster systems of the Milky Way and M87 is that the M87 system is more numerous by nearly a factor of $100$. As a result, in order to use the $148$ simulations of the \texttt{CMC Catalog} to make predictions for M87, we must weight the models according to the M87 cluster properties and then scale up the results. 

In Figure~\ref{fig:CMCmodels}, we plot as gray scatter points cluster mass versus half-light radius for all models in the catalog that survive to an age of $12\,$Gyr, the typical age of globular clusters in both M87 and the Milky Way. In blue and black, we plot the same properties for all clusters observed in the Milky Way \citep{Harris1996} and in M87 \citep{Jordan2009}, respectively.\footnote{In order to compute the present-day cluster mass, we use the integrated $V$-band magnitudes values of \citet{Jordan2009} and assume for simplicity a mass-to-light ratio of 2.}
As evidenced by the plot, the overall distribution in cluster properties between M87 and the Milky Way are remarkably similar, enabling the \texttt{CMC Catalog} models to be used as a representative sample of each cluster population.

To weight the models, we divide the $M_{\rm cl}-r_h$ parameter space of Figure~\ref{fig:CMCmodels} into 12 bins (marked by the solid black lines in the figure). The weight of each model in a given bin is then computed simply as the sum of the masses of all $i$ observed clusters occupying the bin divided by the sum of the masses of all $j$ model clusters in the bin:
\begin{equation}
    \label{eq:weight}
    w = f\times\frac{\sum_i M_{\rm cl,obs}^i}{\sum_j M_{\rm cl,model}^j}.
\end{equation}
The factor $f=5$ incorporates the fact that the total observed globular cluster population of \citet{Jordan2009} encompasses only a fraction of the overall cluster population expected in M87.

\begin{figure*}
    \centering
    \includegraphics[scale=0.7]{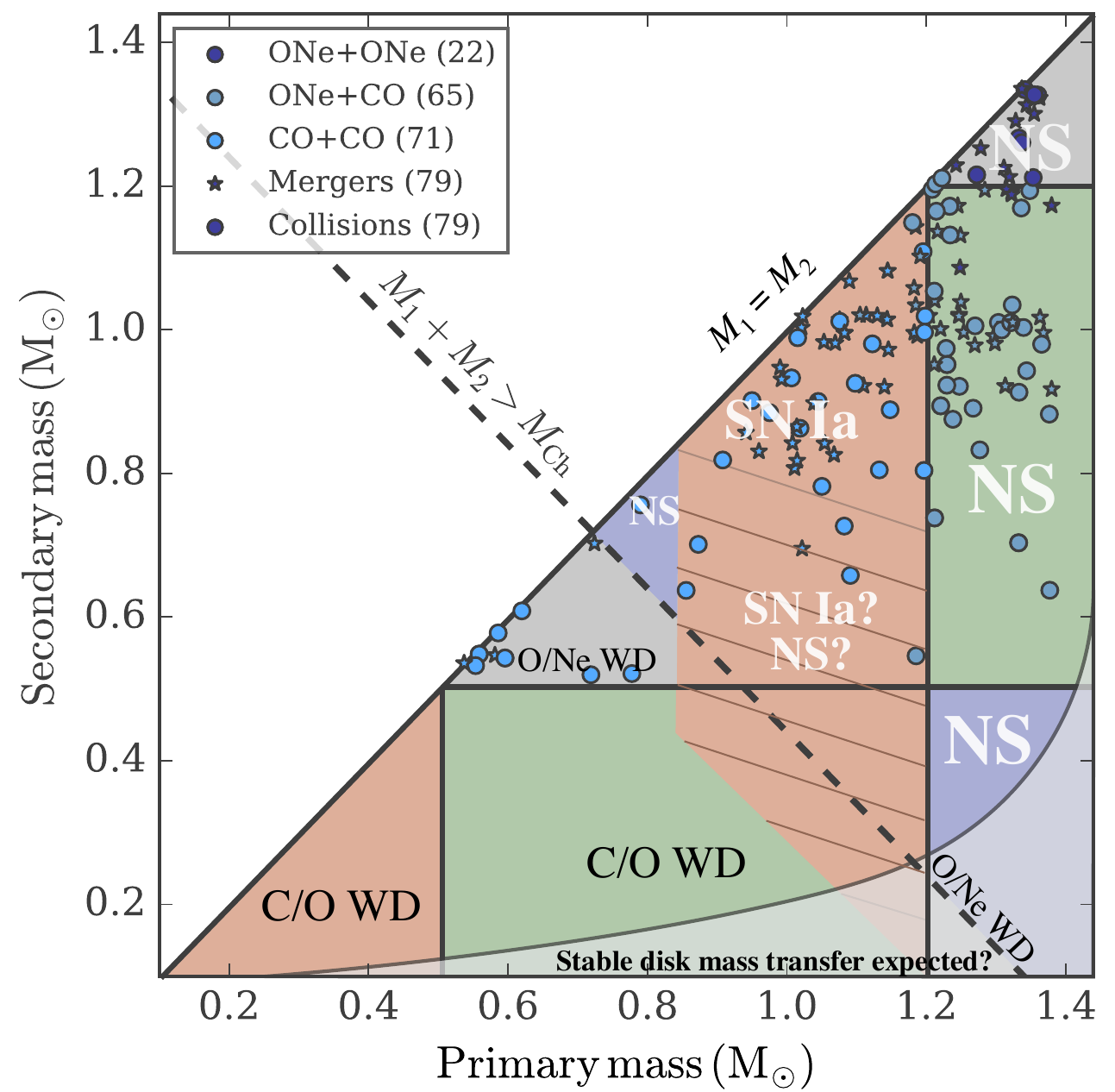}
    \caption{Distribution of white dwarf merger masses for all mergers occurring at late times in our M87 globular cluster sample. Merger outcomes are adapted from \citet{Shen2015}. Solid black lines denote boundaries separating the different white dwarf compositions (also shown as different colored scatter points): helium white dwarfs ($M\lesssim0.5\,M_{\odot}$), carbon-oxygen white dwarfs ($0.5\,M_{\odot} \lesssim M\lesssim1.2\,M_{\odot}$), and oxygen-neon white dwarfs ($M\gtrsim1.2\,M_{\odot}$). The gray shaded region in the bottom-right denotes where an accretion disk forms and stable mass transfer is expected \citep[e.g.,][]{Marsh2004}. As shown, roughly $90\%$ of mergers have mass in excess of $M_{\rm ch}$ (dashed line) and roughly $70\%$ have properties consistent with a neutron star formation outcome.}
    \label{fig:WDmasses}
\end{figure*}

By assigning a weight to each of the white dwarf mergers based on the model in which it occurred following Equation~\ref{eq:weight}, we can estimate the total white dwarf merger rate in the M87 globular clusters. This calculation yields a total of roughly $1.8\times10^5$ white dwarf mergers throughout the full lifetime of globular clusters in M87, roughly $8\times10^4$ of which occur at late times ($t\gtrsim9\,$Gyr) typical of the ages expected for the M87 clusters. This implies a white dwarf merger rate within these old clusters at present day of $\dot{N}_{\rm cl} \approx 2\times10^{-9}\rm{yr}^{-1}$ per cluster. Again using the expression $\tau \approx f_b^{-1} N_{\rm obs}'/\dot{N}_{\rm cl}$ (where $N_{\rm obs}'$ is the number of observable FRB sources per cluster computed in Section~\ref{sec:estimating}) and adopting $f_b=0.3$, this implies a source lifetime of roughly $1.8\times10^6\,$yr, ($9.3\times10^{4},5.5\times10^{6}$) at $90\%$ confidence, consistent with previous estimates from \citet{Kremer2021,Lu2022}.

By further assuming the distribution of M87 cluster properties is representative of the globular cluster systems of other galaxies, we can similarly predict the rate of white dwarf mergers expected in these other galaxies by simply scaling according to the size of the globular cluster population. In column 8 of Table~\ref{table:rates}, we list the inferred rate of white dwarf mergers in each galaxy's globular cluster system.

\subsection{White dwarf merger demographics}

Whether or not a particular massive white dwarf binary merger leads to collapse and formation of a young neutron star capable of emitting FRBs of course depends on the properties of the white dwarfs involved. In Figure \ref{fig:WDmasses}, we show the secondary versus primary mass for all white dwarf mergers identified during a $10\,$Myr time window (representative of the maximum active FRB lifetime expected for a young neutron star formed through such a merger) for an M87 globular cluster sample. Here we have drawn cluster ages randomly from the range $10-13\,$Gyr. The diagonal dashed line marks the $M_1+M_2=M_{\rm Ch}$ boundary. Roughly $90\%$ of the systems identified have mass in excess of $M_{\rm Ch}$, making them viable candidates for collapse \citep[e.g.,][]{NomotoIben1985}. As discussed in \citet{Kremer2021b}, this bias toward massive white dwarf pairs is a direct consequence of mass segregation in globular clusters.

As a background, we show final outcomes expected for the various white dwarf mass and composition combinations, adapted from similar figures in \citet{Dan2014,Shen2015}. This is intended simply to provide visual representation; the exact boundaries of these various outcomes are uncertain. As an example, previous studies have argued pairs of nearly equal mass ($\approx 1\,M_{\odot}$) C/O white dwarfs (occupying the solid red region near the middle of Figure~\ref{fig:WDmasses} are likely to lead to central carbon ignition and a prompt SN Ia \citep[e.g.,][]{Dan2014}. However, this argument is based in part on coarse grids of hydrodynamic simulations and the precise mass ratio that is expected to lead to sufficiently high central temperatures to ignite carbon is uncertain. 

With these caveats regarding the imprecision of the boundaries in mind, it is apparent from Figure~\ref{fig:WDmasses} that a large fraction of the mergers identified in our $N$-body are expected, in principle, to lead to neutron star formation. In particular, 87 out of 158 (roughly $55\%$) of the mergers identified in our M87 sample have at least one O/Ne component, which are considered the likeliest candidates to undergo collapse \citep[e.g.,][]{NomotoKondo1991}.

Collapse to a neutron star is a necessary but not sufficient condition to produce an FRB source. The newly-born neutron star must also have properties (e.g., magnetic field and spin period) that yield appropriate conditions to produce FRBs on a characteristic activity timescale of order $10^5-10^7\,$yr \citep[for discussion of interplay between energetic requirements and timescale requirements, see][]{Kremer2021}. It is not immediately obvious what fraction of white dwarf mergers would lead to such conditions. \citet{Schwab2021} argued that formation of millisecond magnetars with spin periods of $\approx10\,$ms and $B\gtrsim10^{14}\,$G may be possible, depending upon the uncertain details of how much angular momentum is retained through the post-merger luminous giant phase and the magnetic field generated through, e.g., dynamo action in the hot, differentially-rotating merger remnant \citep[e.g.,][]{GarciaBerro2012}. Alternatively, some fraction of these mergers may lead to neutron stars similar to the apparently young pulsars observed at present in a handful of Milky Way globular clusters \citep{Boyles2011} with present-day spin periods of $100-1000\,$ms and inferred $B$-fields of $10^{11}-10^{12}\,$G that have also been linked to formation associated with white dwarf collapse \citep[e.g.,][]{Tauris2013}.

Nonetheless, it appears that the precise fraction of white dwarfs that collapse into neutron stars capable of emitting FRBs can be quite reasonably assumed to be of order unity based on the qualitative features of Figure~\ref{fig:WDmasses} and the predictions from white dwarf merger simulations \citep[e.g.,][]{Schwab2021}. This suggests that massive white dwarf mergers are indeed a viable scenario for producing FRB sources in both M81 and in other nearby galaxies like M87. Importantly, the bursts rates predicted in nearby globular cluster systems in Section~\ref{sec:estimating} is independent of this exact fraction. A lower fraction of white dwarf mergers that lead to FRB sources would simply require a longer active lifetime for bursts to explain the inferred properties of the M81 repeater. 

\section{Testing the white dwarf merger model}
\label{sec:tests}

In this section, we describe a few key features of the white dwarf merger scenario that are in principle testable with a larger population of detected globular cluster FRBs in M87 and other nearby galaxies. 

\subsection{Offset from host cluster center}

The median radial position at which the late-time white dwarf mergers identified in our models occur within their host cluster is roughly $0.04\,$pc -- this clear preference for occurring near their host cluster's center is a result of mass segregation of the white dwarf progenitors \citep{Kremer2021b}. Although the mergers themselves are expected to be centrally-concentrated, the FRB-emitting neutron stars subsequently formed are expected to be offset from their host clusters' centers as a result of velocity kicks imparted during the merger process and/or during the supernova associated with neutron star formation. In this case, the observed offset of FRBs may be used to constrain the uncertain magnitude of these kicks.

The M81 FRB is observed to be offset by roughly $2\,$pc (two-dimensional projection) from its host cluster's center \citep{Kirsten2022}. For simplicity, we can adopt a Plummer potential for the cluster, \citep[e.g.,][]{HeggieHut2003}

\begin{equation}
    \phi(r)=-\frac{GM_{\rm cl}}{a}\Bigg(1+\frac{r^2}{a^2}\Bigg)^{-1/2}
\end{equation}
with $M_{\rm cl}=5.8\times10^5\,M_{\odot}$ and $a\approx0.77r_h\approx2.84\,$pc for the Plummer scale radius (assuming $r_h=3.7\,$pc) as inferred from observations \citep{Kirsten2022}. This implies a central velocity dispersion $\sigma_v=\sqrt{GM_{\rm cl}/6a}\approx12\,\rm{km\,s}^{-1}$. The velocity, $v_i$, of the neutron star immediately after the kick is given by 

\begin{equation}
    \label{eq:v_i}
    v_i^2=v_k^2+v_{\rm th}^2+2v_k v_{\rm th}\cos\theta ,
\end{equation}
where $v_k$ is the kick velocity magnitude, $v_{\rm th}\approx12\,\rm{km\,s}^{-1}$ is the magnitude of the ``thermal'' velocity associated with the cluster's central velocity dispersion, and $\theta$ is the angle between $\bf{v}_{\it k}$ and $\bf{v}_{\rm th}$. Assuming conservation of energy along a radial orbit, we have 
\begin{equation}
    \label{eq:energy}
    \mathcal{E} = \frac{1}{2}v_i^2 + \phi(0) = \frac{1}{2}v(r)^2 + \phi(r)=\phi(r_{\rm max}).
\end{equation}
In this case, the minimum initial kick velocity required to achieve $r_{\rm max}=2\,$pc is

\begin{equation}
    \label{eq:min_kick}
    v_{\rm k,min} = \sqrt{ \frac{2GM_{\rm cl}}{a} \Bigg[1-\Bigg(1+\frac{r_{\rm max}^2}{a^2}\Bigg)^{-1/2} \Bigg] - v_{\rm th}^2 \sin^2\theta}  - v_{\rm th} \cos \theta.
\end{equation}
Assuming $\theta=\pi/2$ (the optimal case where the $\bf{v}_{\rm k}$ is aligned with $\bf{v}_{\rm th}$) and plugging in relevant values for other parameters, we obtain $v_{\rm k,min}\approx13\,\rm{km\,s}^{-1}$.\footnote{If, due to projection effects, the true three-dimensional radial offset of the source is larger than the $2\,$pc two-dimensional offset, this would require a larger kick. For example, a projection of $45^\circ$ would imply a true radial offset of roughly $2.8\,$pc, requiring $v_{\rm{k,min}}\approx19\,\rm{km\,s}^{-1}.$ Thus, the zero projection case gives a true minimum kick value.} The one-way travel time from $r=0$ to an arbitrary $r$ is computed as \begin{equation}
    \label{eq:t_r}
    t(r)=\int_0^{r}\frac{dr'}{ \sqrt{2[\mathcal{E}-\phi(r')]}}.
\end{equation}
We define $t_0$ as the travel time from $r=0$ to $r_{\rm obs}=2\,$pc and $t_{\rm apo}$ as the travel time to reach the apocenter distance of the given radial orbit, $r=r_{\rm max}$. 

Accounting for possibility of repeated orbits, we can then write the allowed age of the neutron star presently observed at $r=2\,$pc as

\begin{equation}
    \label{eq:age}
    \tau = (-1)^{n+1} t_0 + 2 \left\lfloor \frac{n}{2} \right\rfloor t_{\rm apo}, \,\,\, n=1,2,3,...
\end{equation}
where the function $\lfloor n/2 \rfloor$ selects the largest integer $\leq n/2$.

\begin{figure}
    \centering
    \includegraphics[scale=0.8]{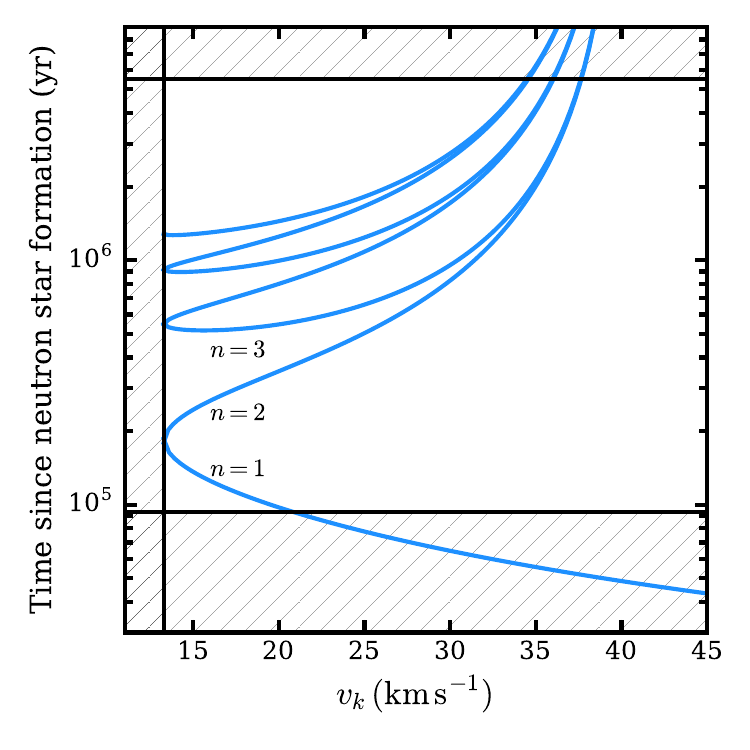}
    \caption{Blue curves show allowed combinations of neutron star age (vertical axis) and velocity kick values (horizontal axis) for a few different values of $n$ (Equation~\ref{eq:age}). Hatched regions are ruled out, as discussed in the text. From this plot, we infer an allowed range of kick velocities of roughly $13-38\,\rm{km\,s}^{-1}$, plausibly consistent with natal kick velocities expected from ultra-stripped SNe.}
    \label{fig:kick}
\end{figure}

From Equations~(\ref{eq:v_i}), (\ref{eq:energy}), (\ref{eq:t_r}), and (\ref{eq:age}), we can compute age versus $v_k$ curves for different values of $n$. We show these results as blue curves in Figure~\ref{fig:kick}. Again, we have adopted $\theta=\pi/2$. The vertical solid black line denotes the $v_{\rm k,min}=13\,\rm{km\,s}^{-1}$ boundary. $v_k$ values to the left of this boundary are ruled out. As horizontal solid black lines, we denote the $90\%$ age values of $9.3\times10^4\,$yr and $5.5\times10^6\,$yr inferred from the M81 event, as discussed in Section~\ref{sec:estimating}. Age values above and below these boundaries are also ruled out.

The maximum allowed kick velocity value can be identified as the maximum crossing point of a blue curve with any of the forbidden hatched gray regions. We identify $v_{\rm k,max}\approx 38\,\rm{km\,s}^{-1}$ from the crossing points of the $n=2,3$ curves. Thus, the kick is constrained to lie within the range $\approx 13-38\,\rm{km\,s}^{-1}$. We now discuss possible origins of this kick.

During the merger, a fraction of mass is likely unbound from the system dynamically. The exact amount of mass loss depends on the mass ratio, but hydrodynamic simulations suggest ejecta values of roughly $10^{-3}\,M_{\odot}$ are typical \citep{Dan2014}. This ejecta has a characteristic value of order the escape speed of the white dwarf prior to disruption, $v_{\rm ej}\sim10^4\,\rm{km\,s}^{-1}$. Conservation of linear momentum implies the remaining white dwarf merger remnant receives an impulsive kick of characteristic value

\begin{equation}
    \label{eq:merger_kick}
    v_{\rm k}^{\rm merger} \approx 5\,\rm{km\,s}^{-1} \Bigg(\frac{\it{M}_{\rm ej}}{10^{-3}\,\it{M}_{\odot}}\Bigg) \Bigg(\frac{\it{M}_{\rm tot}}{2\,\it{M}_{\odot}}\Bigg)^{-1} \Bigg(\frac{\it{v}_{\rm ej}}{10^4\,\rm{km\,s^{-1}}}\Bigg)^{-1}
\end{equation}
where we have assumed the optimal case where the momentum is imparted along a single direction. Since the characteristic value inferred is less than the minimum allowed kick value of roughly $13\,\rm{km\,s}^{-1}$, we conclude a merger kick alone is likely insufficient.

On timescales of roughly $10\,$kyr after the white dwarf merger, the merger remnant is expected to undergo an \textcolor{magenta}{\bf electron-capture} or iron core-collapse supernova explosion (as a Type Ic supernova, since the envelope -- consisting of the previously disrupted white dwarf -- is hydrogen/helium-poor), leaving behind a neutron star \citep[e.g.,][]{Shen2015, Schwab2016,Schwab2021}. During the supernova, a momentum kick of velocity, $v_{\rm k}^{\rm SN}$, is expected to be imparted to the newborn neutron star associated with explosion asymmetries \citep[e.g.,][]{BlondinMezzacappa2006}. Neutron star natal kick velocities associated with standard iron CCSNe are expected to be of order a few $100\,\rm{km\,s}^{-1}$ \citep[e.g.,][]{Hobbs2005}. However, standard iron CCSN physics may not necessarily apply to the collapse of white dwarf merger remnants. For instance, \citet{Tauris2015} demonstrated that, for ultra-stripped SNe with very low ejecta mass compared to standard SNe, the explosion may result in a relatively weak impulse on the neutron star. Previous studies have demonstrated kick velocities associated with ultra-stripped SNe of $\lesssim50\,\rm{km\,s}^{-1}$ \citep[e.g.,][]{Tauris2015,Suwa2015, Tauris2017,Janka2017,De2018}, inferred in part through the measured orbital properties of known double neutron star systems \citep[e.g.,][]{Tauris2006,Ferdman2013}. Interestingly, this is consistent with our inferred range of allowed velocities.

Although this discussion is intended only as a rough estimate of the allowed kick velocity value (for example, we have not incorporated uncertainties in the underlying white dwarf merger rate or in the potential profile of the host cluster), we tentatively suggest these arguments as evidence for an ultra-stripped-SN-like natal kick connected with the formation of the M81 FRB source. In principle, with a larger sample of in-cluster FRBs with observed offsets and host cluster properties, neutron star natal kick velocities may be further constrained in this manner.

Besides white dwarf merger-induced collapse, young neutron stars similarly capable of powering FRBs may also form through accretion-induced collapse \citep[AIC;][]{Kirsten2022,Kremer2021}. In this scenario, a massive ONeMg white dwarf accretes sufficient material from a binary companion to reach the Chandrashekhar limit, triggering runaway electron capture and collapse to a neutron star \citep[e.g.,][]{NomotoKondo1991,Tauris2013,Schwab2015}. Neutron star formation via AIC is expected to also be accompanied by a kick of order $10\,\rm{km\,s}^{-1}$ \citep[e.g.,][]{Podsiadlowski2004,Kitaura2006}, of comparable value to that expected from the white dwarf merger collapse scenario. Thus, the AIC formation scenario may also lead to an off-center FRB source. For the same mass segregation arguments that apply for white dwarf mergers, AIC events in old core-collapsed globular clusters are expected to preferentially occur near their host cluster's center. However, one key difference in the AIC scenario is that the newly born neutron star may still reside in a binary system (assuming the natal kick does not disrupt the binary). In this case, the center-of-mass kick is reduced by a factor of $\sim2$ compared to that received by the neutron star. Taking this into account, if the M81 FRB source formed through AIC and has a binary companion, the natal kick velocity required to explain the observed cluster offset would be larger than in the white merger scenario, roughly $25-75\,\rm{km\,s}^{-1}$. This range of values is still physically plausible based on the expected central escape speed of the host cluster and expectations for kicks arising from electron-capture SNe. In this case, given that the expected formation locations and expected kick velocities appear to be comparable for the merger scenario and AIC scenario, it is not clear whether an observed FRB offset may favor one scenario versus the other. Lastly, we note that if some fraction of globular cluster FRB sources do have binary companions, this may produce interesting observational consequences, for example periodicity in burst repetition \citep[e.g.,][]{Lyutikov2020}, persistent X-ray emission \citep[from possible subsequent accretion through Roche lobe overflow; e.g.,][]{Tauris2013}, or time-dependent dispersion measure/rotation measure. We reserve for future work consideration of these possible complexities.

\subsection{Intercluster FRB repeaters from ejected white dwarf binaries}

\begin{figure}
    \centering
    \includegraphics[scale=0.75]{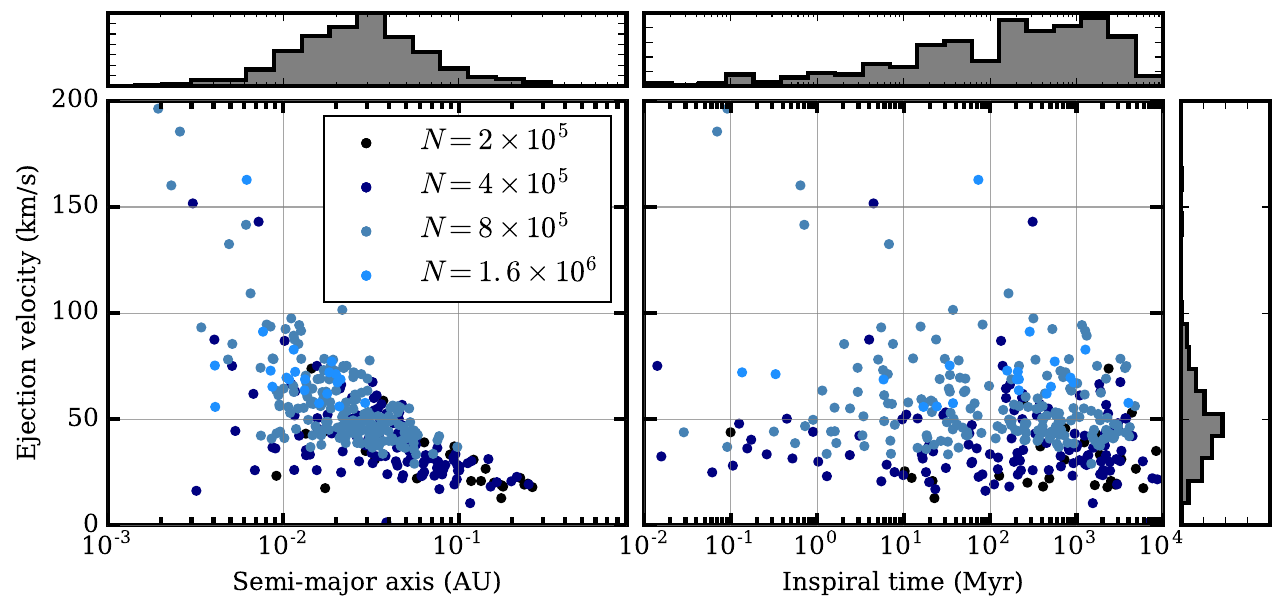}
    \caption{\textit{Left panel:} Velocity versus semi-major axis at time of ejection from host cluster for all ejected white dwarf binary mergers in the \texttt{CMC Catalog} simulations. \textit{Right panel:} Velocity versus inspiral time at moment of ejection. Different colors denote different initial $N$ for the models.}
    \label{fig:escape_velocities}
\end{figure}

As a natural consequence of the dynamical interactions in globular clusters that lead to the formation of white dwarf binaries and subsequent mergers, a population of compact white dwarf binaries are expected to be dynamically ejected from their host cluster \citep{Kremer2021b}. Simple energy arguments suggest that the characteristic dynamical recoil velocity of a binary in a cluster following a resonant encounter with another object scales with its orbital velocity, $v_{\rm recoil} \sim v_{\rm orb} \sim \sqrt{GM/a}$ \citep[e.g.,][]{HeggieHut2003}. As a binary hardens, it receives increasingly large recoil kicks, which, for sufficiently compact binaries, can exceed the central escape speed, $v_{\rm esc}$ of the host cluster. The characteristic orbital separation at which a $1\,M_{\odot}+1\,M_{\odot}$ white dwarf binary is ejected is $a_{\rm ej} \approx 0.02\,\rm{AU} (\it{m}/M_{\odot})(v_{\rm esc}/\rm{40\,km\,s}^{-1})^{-2}$ \citep{Kremer2021b}. Assuming an eccentricity $e=0.9$ (typical of dynamically-formed binaries), the gravitational wave inspiral time of such a binary after ejection is $t_{\rm insp}\sim 100\,\rm{Myr}(\it{a}/\rm{0.02\,AU})^4(\it{m}/M_{\odot})^{-\rm{3}}(\rm{1}-\it{e}^2)^{\rm{7/2}}$ \citep[e.g.,][]{PetersMathews1963}. In this case, these ejected binaries will merge outside of their host clusters, potentially creating young neutron stars similarly capable of producing (repeating) FRB sources in the halo of their host galaxies with properties similar to the M81 source.

In our \texttt{CMC Catalog} models, we identify 357 total ejected super-Chandrasekhar white dwarf binary mergers at late times. We plot the properties at time of ejection for all mergers in Figure~\ref{fig:escape_velocities}. The left panel shows the velocity versus binary semi-major axis at time of ejection. This panel illustrates the anticipated anti-correlation between $v_{\rm recoil}$ and $a$, as described in the previous paragraph. The right panel shows velocity versus inspiral time. Here, the anti-correlation of the left panel becomes ``smeared” out due to the random eccentricities of the binaries at time of ejection and the steep  dependence of $t_{\rm insp}$ on eccentricity. Taking the $v$ and $t_{\rm insp}$ values shown in Figure~\ref{fig:escape_velocities}, we can integrate the orbits of these binaries in an M87-like potential to determine the typical distance travelled from their host at time of merger. We adopt an NFW potential for M87 with $M_{\rm vir}=10^{14.4}\,M_{\odot}$ and scale radius $r_s=448\,$pc \citep{OldhamAuger2016}. We draw initial radial positions from the observed cluster radial distribution over the range $0-40\,$kpc (the field of view for FAST; see Figure~\ref{fig:radial_dist}) and draw a random polar angle to determine $z$ and 2D radial position. We draw random angles $\theta$ and $\phi$ to convert $v$ values shown in Figure~\ref{fig:escape_velocities} into an initial 3D velocity vector. With these initial conditions, we then integrate the orbits for time $t_{\rm insp}$ using the \texttt{galpy} package \citep{Bovy2015_galpy}. We find the median separation from their original host cluster position of these binaries at time of neutron star formation is about $13\,$kpc. Roughly $85\%$ of binaries have a separation of at least $1\,$kpc (the exceptions being systems with very short inspiral times; right panel of Figure~\ref{fig:escape_velocities}). Thus, a large cluster offset is expected for the majority of these ejected mergers.

The results from the \texttt{CMC Catalog} imply roughly $3\times10^4$ ejected mergers in an M87 globular cluster sample, using the same weighting scheme described in Section~\ref{sec:WD}. Assuming, as with the in-cluster mergers, that each of these ejected mergers leads to collapse and formation of an FRB source, this implies roughly one intercluster FRB for every three in-cluster FRBs detected in the M87 globular cluster system. This implies, based on the burst detection rates estimated in Section \ref{sec:estimating}, roughly $7.35^{+14.72}_{-6.98}$ (for $\alpha=-2.4$) or $2.49^{+4.98}_{-2.37}$ (for $\alpha=-2$) intercluster FRBs are expected per hour of observation of M87 by FAST. The detection (or lack thereof given enough observing time) of such intercluster FRBs may provide evidence for or against the white merger dynamical formation scenario outlined here. Finally, since only a single in-cluster FRB source is predicted in Cen A (Table~\ref{table:rates}), it is unlikely any intercluster FRB sources are currently detectable in Cen A.

\subsection{Red versus blue cluster populations}

It is well-established that the color distributions of extragalactic globular cluster systems are bimodal arising from two distinct cluster populations: metal-rich red clusters and metal-poor blue clusters \citep[e.g.,][]{BrodieStrader2006}. Although the exact origin of this bimodal distribution remains uncertain \citep[for possible formation scenarios, see, e.g.,][]{AshmanZepf1992,WhitmoreSchweizer1995,Forbes1997,Cote1998,Harris1999}, it seems clear that red clusters are, in general, relatively young compared to their blue counterparts due to the well-known age-metallicity degeneracy \citep[e.g.,][]{Worthey1994}. As an example, in M87, \citet{Kundu1999} estimated the red clusters were born in burst of star/cluster formation $3-6\,$Gyr after the blue clusters. On the other hand, \citet{Jordan2002} found the red and blue subpopulations to be coeval, estimating a much smaller typical age difference of $0.2\pm 2\,$Gyr.

As summarized in Section \ref{sec:WD}, clusters evolve toward core collapse on a timescale of order their two-body relaxation time. In this case, for two clusters of similar mass and size at birth, the older cluster is more likely to have reached core collapse by the present day. Thus, on average one may expect a larger fraction of blue clusters to have reached core-collapse compared to red clusters. Since, as discussed in Section \ref{sec:WD}, we argue core-collapsed clusters yield by far the highest rate of white dwarf mergers, one would naturally expect blue clusters, on average, to host higher rates of FRBs. In M87, roughly 70\% (30\%) of globular clusters observed are categorized as blue (red) \citep[e.g.,][]{Strader2011}. Assuming a typical age of $9\,$Gyr for red clusters \citep{Kundu1999}, it is likely that relatively few red clusters have reached core collapse and, as a consequence, the present-day white dwarf merger rate in red clusters is expected to be several orders of magnitude lower than the rate in blue clusters. As a result, the specific abundance (number per cluster) of FRBs in red clusters should be several orders lower than that in blue clusters. In this case, the detection of even a single FRB in a red globular cluster may hint at formation mechanisms other than the white dwarf merger scenario. Note that the globular cluster host of the M81 FRB has metallicity $[\rm{Fe/H}]=-1.83^{+0.86}_{-0.87}$ \citep{Kirsten2022}, the median value of which would clearly categorize this as a blue cluster, similar to the blue population observed in M87.

\subsection{Cosmological rate distribution}

\begin{figure}
    \centering
    \includegraphics[scale=0.75]{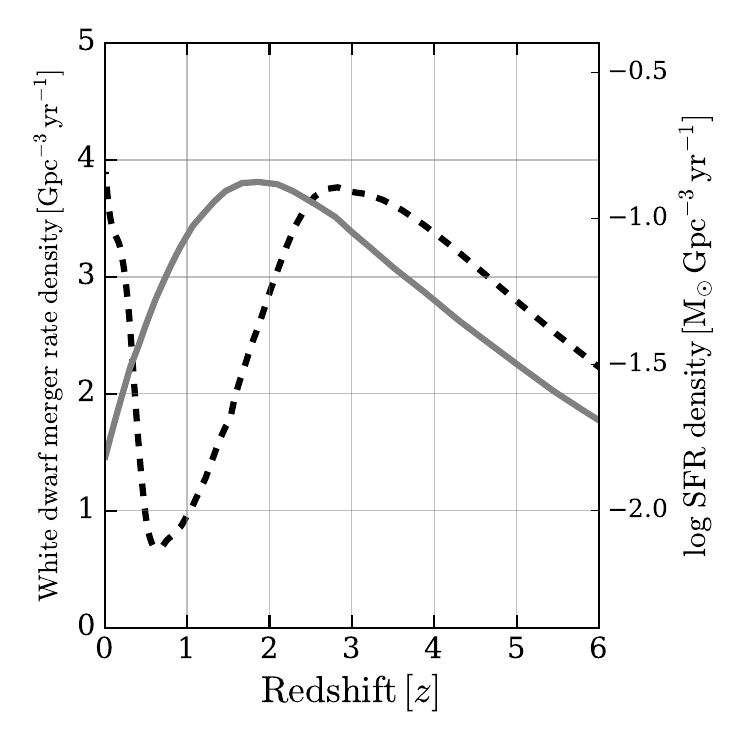}
    \caption{Globular cluster white dwarf merger rate density (dashed line) versus redshift. The peak at $z\approx 0$ comes from the increase in white dwarf mergers occurring in core-collapsed globular clusters. The peak at $z\approx3$ is associated with primordial white dwarf binary mergers occurring at early time which track roughly the cluster formation times. For reference, we show (on the right-hand vertical axis) the star formation rate density (solid gray line) versus redshift from \citet{MadauDickinson2014}.}
    \label{fig:rate}
\end{figure}

Last, we comment on the expected cosmological rate of FRBs formed through the white dwarf merger scenario. First, we compute the volumetric rate of white dwarf mergers in clusters versus redshift. We follow the same method outlined in \citet{kremer2020modeling} for the case of binary black hole mergers in clusters. The (comoving) volumetic rate at a given redshift is computed as $\mathcal{R}(z)=\rho_{\rm GC}dN(z)/dt$, where $\rho_{\rm GC}$ is the volumetric number density of globular clusters \citep[we adopt a constant value $\rho_{\rm GC}=2.31\,\rm{Mpc}^{-3}$; for discussion, see][]{Rodriguez2015} and $dN/dt$ is the number of white dwarf mergers per unit time at a given redshift. To compute $dN/dt$, we: (1) generate a complete list of all white dwarf mergers occurring in our $N$-body simulations; (2) weight each cluster model according to the weighting scheme described in Section~\ref{sec:WD}; (3) draw a random age for the host cluster from which each merger originated. For each merger, we draw 10 independent cluster ages from the metallicity-dependent age distributions of \citet{ElBadry2019}; (4) count up the total number of mergers per unit time ($dN/dt$) by dividing into a discrete set of redshift bins; (5) divide this rate by a factor of 10 to account for the oversampling in age draws; (6) divide by an additional factor of $2000$ to scale down the (weighted) sample of models to a single typical cluster.

In Figure~\ref{fig:rate}, we show the results of this calculation as a dashed black curve. Assuming all of these white dwarf mergers lead to young neutron star formation, this curve can be interpreted as the formation rate density of FRB sources in globular clusters versus redshift. As shown, the volumetric rate curve features two peaks: the first, at roughly $z\approx3$ is associated with the short-delay-time ($\lesssim100\,$Myr) mergers occurring through evolution of primordial cluster binaries. The location (and width) of this peak is sensitive primarily to the peak and duration of the assumed cluster formation history. The second peak at $z\approx0$ arises from the increase in white dwarf mergers at late times as the clusters evolve toward core collapse and attain dense subsystems of white dwarfs in their centers. The height of this $z\approx0$ peak is determined by the fraction of clusters in the local universe that have reached core-collapse and the width is determined by the typical age at which the core-collapsed clusters reached a core-collapsed state.

\begin{figure}
    \centering
    \includegraphics[scale=0.8]{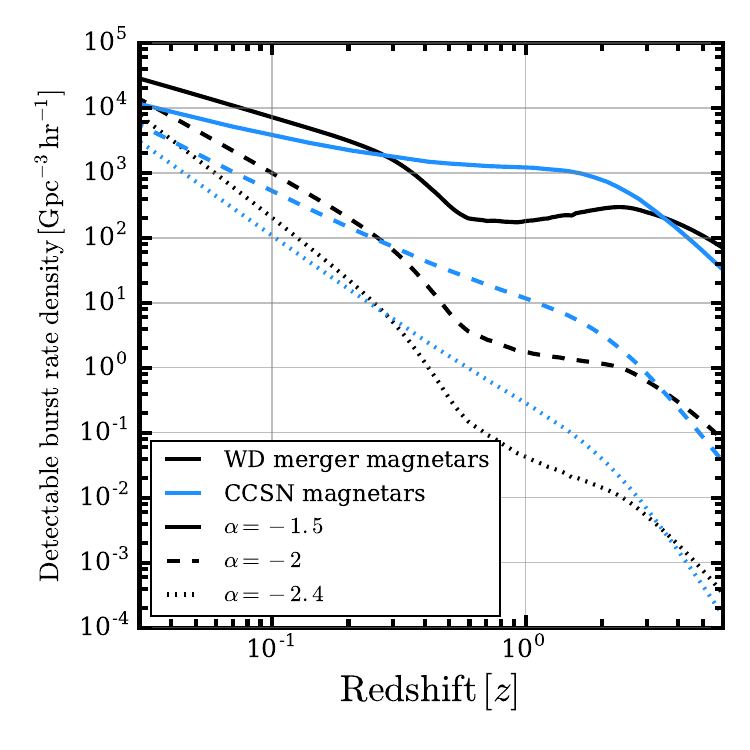}
    \caption{Detection rate density versus redshift for magnetars formed through white dwarf mergers (tracking the dashed curve in Figure~\ref{fig:rate}) and magnetars formed through CCSNe (tracking the solid gray curve in Figure~\ref{fig:rate}).}
    \label{fig:rate_detect}
\end{figure}

We show as a solid gray curve on a secondary y-axis the star formation rate density of \citet{MadauDickinson2014}. The formation rate of magnetars through CCSNe is expected to trace this curve, normalized by the fact that roughly $0.0068$ core-collapse events are expected per $M_{\odot}$ of stars formed \citep[e.g.,][]{MadauDickinson2014} and of these core collapse events, roughly $10\%$ are expected to lead to magnetar formation \citep[e.g.,][]{KaspiBeloborodov2017}.

In Figure~\ref{fig:rate_detect}, we show the detection rate density versus redshift. The value is computed at a given redshift as $\mathcal{R}(z)\times \tau \times R$. Here $\mathcal{R}(z)$ is the source formation rate density as shown in Figure~\ref{fig:rate} (for the CCSN magnetars, we assume $0.0068\times0.1$ magnetars formed per $M_{\odot}$, as discussed above), $\tau$ is the active FRB lifetime of each source \citep[we adopt $\tau=10^6\,$yr for the white dwarf merger magnetars as discussed in Section \ref{sec:estimating} and adopt $\tau=100\,$yr for CCSN magnetars, consistent with the characteristic ages of Galactic magnetars;][]{KaspiBeloborodov2017}, and $R$ is the burst rate described by Equation~(\ref{eq:energy}). Again, we adopt three values for the uncertain $\alpha$ power-law index, shown as solid, dashed, and dotted curves in the figure. Although quite speculative at present, in principle, with a large enough sample of FRB detections that can be identified with one population or the other, these detection rate densities may be constrained observationally. 

\section{Summary and Conclusions}
\label{sec:discussion}

The detection of a repeating FRB localized to an old globular cluster in the halo of M81 challenges our understanding of both FRB physics and globular cluster dynamics. Based on the presence of the M81 FRB source, we estimate the number of similar FRB sources detectable in the globular cluster systems of nearby galaxies, utilizing the ``Globular Cluster Systems of Galaxies Catalog" presented in \citet{Harris2013}. Next, using a large suite of $N$-body cluster simulations, we go onto explore massive white dwarf mergers as a possible formation mechanism for these FRB sources and discuss several key features and testable predictions of this scenario. Our main results and conclusions are:
\begin{enumerate}
    \item We predict M87 -- known to host in excess of $10^4$ globular clusters -- contains the most active FRB sources of all nearby galaxies at present day, up to $\mathcal{O}(10)$ sources.
    \item By scaling to the detected burst rate of the M81 FRB source and incorporating the uncertain burst energy distribution of such sources, we estimate the detectable burst rate in these globular cluster systems. We identify M87 and Cen A as the most promising targets for radio telescopes such as FAST and MeerKat, respectively. We predict a dedicated radio survey of M87 of duration roughly $30\,$hr ($15\,$hr) by FAST (MeerKat) has a $90\%$ chance of detecting at least one globular cluster FRB, even for our most pessimistic assumed burst energy distribution.
    \item Young highly-magnetic neutron stars formed through collapse following massive white dwarf mergers have been identified as a promising formation scenario for FRBs like the M81 source. Motivated by this hypothesis, we use a large suite of $N$-body globular cluster simulations to predict the merger rate of massive white dwarf binaries in the globular cluster systems of various galaxies.
    \item We explore the properties of the white dwarf mergers occurring in our $N$-body simulations and establish that: (i) the vast majority (roughly $90\%$) have total mass in excess of the Chandrasehkar limit and (ii) over half have mass ratios and white dwarf compositions consistent with those expected to lead to collapse to a neutron star based on expectations inferred from previous simulations of white dwarf mergers \citep[e.g.,][]{Dan2014,Shen2015,Schwab2016,Schwab2021}. In this case, it seems quite plausible that a fraction of order unity of these mergers may indeed lead to FRB sources.
    \item Using the M81 FRB as a test case, we described how the observed offset of FRBs within their host globular clusters may be used to constrain supernova natal kicks associated with the final evolutionary stages of these massive white dwarf merger remnants. For the M81 FRB, we constrain a supernova kick velocity of $\approx13-38\,\rm{km\,s}^{-1}$, remarkably consistent with predictions of kick velocities arising from ultra-stripped supernovae. Alternatively, a young neutron star formed via accretion-induced collapse may receive a comparable recoil kick at birth and thus produce FRB sources with comparable cluster offsets. In principle, an ensemble of FRB sources detected in globular clusters with similarly-constrained cluster offsets may be used to place further constraints on such supernova kicks which, at present, lack robust observational constraints.
    \item As described in \citet{Kremer2021b}, a consequence of the dynamical formation of white dwarf binary mergers in globular clusters is the dynamical ejection of a subset of compact white dwarf binaries from their hosts. These binaries go onto merge (and presumably collapse to form neutron stars) in their host galaxy's halo roughly $100\,$Myr (on average) after ejection from their host clusters. These ejected post-merger neutron stars may be detectable as host-less ``intercluster'' FRB sources in the halos of their host galaxies. We predict that roughly one intercluster FRB source should be present for every three in-cluster sources. This implies of order one to a few intercluster sources in the halo of M87 at present.
    \item Finally, we described the cosmological evolution of FRB rates from magnetars formed in white dwarf mergers in clusters and demonstrated this rate may be distinguishable from that of FRBs arising from magnetars formed through core-collapse supernovae.  
\end{enumerate}

The exact details behind the mechanism of FRB sources remain mysterious, due in large part to the cosmological distances at which the vast majority of FRBs are observed. The detection of even a small number of additional FRBs like the M81 source in nearby galaxies that could be localized to specific globular clusters would be pivotal to the FRB field. Our results suggest such detections are not only possible, but likely, motivating targeted radio surveys of the globular cluster systems of local galaxies. 

\acknowledgements
KK is supported by an NSF Astronomy and Astrophysics Postdoctoral Fellowship under award AST-2001751. We thank the anonymous referee for a careful review and many helpful comments that improved the manuscript.

\bibliographystyle{aasjournal}
\bibliography{mybib.bib}

\end{document}